**Title:** Realizing Nonreciprocal Linear Dichroism and Emission from Simple Media

**Authors:** Thomas J. Ugras[1,2], Daniel J. Gracias[3], Oriol Arteaga[4], Richard D. Robinson[2,3]*

**Affiliations:**

[1] School of Applied and Engineering Physics, Cornell University, Ithaca, NY, USA.

[2] Kavli Institute at Cornell for Nanoscale Science, Ithaca, NY, USA.

[3] Department of Materials Science and Engineering, Cornell University, Ithaca, NY, USA.

[4] Departament de Física Aplicada, Universitat de Barcelona, IN2UB, Barcelona, Spain.

*corresponding author: rdr82@cornell.edu

**ORCiD:**

Thomas J. Ugras: 0000-0002-5931-3031

Daniel J. Gracias: 0009-0007-2358-831X

Oriol Arteaga: 0000-0001-9015-0237

Richard D. Robinson: 0000-0002-0385-2925

**Abstract:** Reciprocity – the principle that a system response is identical in the forward path compared to the backward path – is a fundamental concept across physics, from electrical circuits and optics to acoustics and heat conduction. Nonreciprocity arises when this symmetry is broken, enabling directional-dependent behavior. In photonic systems, nonreciprocity allows control over the propagation of electromagnetic waves, essential for isolators and circulators. But achieving optical nonreciprocity typically requires complex metamaterials, exotic media, or strong external fields. Even materials with asymmetric transmission that preserve time-reversal symmetry often rely on intricate fabrication processes. Because of this, researchers have historically overlooked the possibility that readily available materials could support nonreciprocal optical behavior, assuming that conventional systems lack the ability to produce nonreciprocal behavior. In this work, we challenge that assumption by revisiting the light-matter interactions of chiroptic and linearly anisotropic media. Through Stokes-Mueller formalism we derive a simple analytical expression that predicts a pathway to nonreciprocal absorption and emission of orthogonal linear polarizations. We test this idea experimentally using solution-processed films of CdS, CdSe, and CdTe magic-size clusters that possess commensurate circular dichroism (CD) and linear dichroism (LD) values and find that they can support this effect, engineering films that exhibit nonreciprocal absorption and emission of linearly polarized light. Based on the derived expressions and experiments, several design rules are presented. Our findings reveal that nonreciprocal linear dichroism and emission can be achieved in readily processable, macroscopically symmetric materials by harnessing chiral-linear optical interference. This work opens new opportunities for scalable, polarization-based photonic control for direction-dependent optical routing, optical logic, and polarization-multiplexed information encoding.

**Main Text:**
Reciprocity, the principle that light traveling along a forward path experiences the same response when retracing that path in reverse, is a foundational symmetry in optics. This behavior is common, and assumed, in everyday objects like polarized sunglasses, which function identically regardless of which side faces the light. And the same symmetry is embedded in most linear, passive, time-invariant optical systems, like lenses, mirrors, birefringent crystals, and diffraction gratings.

Breaking this symmetry opens the door to fundamentally new physical principles and enables advanced photonic technologies such as isolators, circulators, and optical diodes, devices that allow light to propagate in only one direction. Typical optical isolators, for instance, primarily rely on magneto-optic materials, such as yttrium iron garnet (YIG), which exploit Faraday rotation under an external magnetic field to achieve nonreciprocity (*1*).

Yet, breaking reciprocity is far from trivial. And even realizing the simpler forms of nonreciprocity, such as asymmetric optical transmission and polarimetric nonreciprocity, that do not require violation of Lorentz reciprocity, are difficult, requiring complex, lithographic structures, periodic media, and exotic materials (*2-6*). Researchers have explored alternative approaches using nonlinear optics and time-modulated photonic systems, but these remain largely at the experimental or proof-of-concept stage and have yet to match the robustness and integration challenges of magneto-optic solutions.

In recent years, researchers have made advances on this issue by leveraging misaligned optical properties to create nonreciprocal circularly dichroic (NRCD) films (*7-11*). Analogous progress has been made in emission leveraging the same optical anisotropy, where films have been shown to emit orthogonal circular polarizations through opposite faces of the sample (*12-14*).

Despite this progress, no analogous mechanism has been theoretically proposed or experimentally demonstrated for linear polarization. Yet, arguably, nonreciprocal effects involving linear polarization may play an even greater role in emerging technologies, as they underpin many fundamental light-matter interactions. In particular linearly dichroic effects are often orders of magnitude stronger than their circular counterparts, making them especially relevant for practical applications.

In this work, we revisit the foundational Stokes-Mueller formalism describing light-matter interactions to derive a relationship between optical properties in materials that enable nonreciprocal linear dichroism (NRLD), the preferential absorption of orthogonal linear polarizations based on the direction of incidence in passive films.

To experimentally test our theory, we leverage films composed of CdS magic-size clusters (MSCs) and find that they can exhibit NRLD, flipping the sign of LD based on the direction of propagation. Further, we model the spectral response and find good agreement between the measurements and a model. We generalize and substantiate these findings by demonstrating the effect in films made of CdSe and CdTe, and by achieving agreement between measured spectra and models. The discovery of NRLD is extended to nonreciprocal emission of linear polarizations, which we derive from theory and demonstrate experimentally with the CdS MSC films. Finally, we draw parallels and differences between NRLD and natural LD and between NRLD and NRCD.

**Concept**

To investigate how to achieve nonreciprocity from the interactions of a material's optical properties, we derived expressions for measured circular dichroism (CD) and LD signals through Stokes-Mueller calculus (*15-17*). Differential Stokes-Mueller matrix formalism describes the evolution of polarized light as it propagates through a homogeneous medium (*18*). Following this method, the measured values for the CD and LD signals are described by the following equations (full derivation in SM):

$$CD_{measured} \approx CD + \frac{1}{2}(LBLD` - LDLB`) \quad (1a)$$

$$LD_{measured} \approx (LD` \sin 2\theta - LD \cos 2\theta) + \frac{1}{2}\big((CBLD - CDLB)\sin 2\theta + (CBLD` - CDLB`)\cos 2\theta\big) \quad (2a)$$

where LD is linear dichroism, LB is linear birefringence, LD` and LB` are their prime counterparts measured at ±45°, CD is circular dichroism, CB is circular birefringence, and $\theta$ is the orientation of the film with respect to the light propagation direction. The terms for these expressions can be re-grouped based on the contributions as

$$CD_{measured} \approx CD_{chiroptic} + CD_{LDLB} \quad (1b)$$
$$LD_{measured} \approx LD_{natural} + LD_{CDLB} \quad (2b)$$

Where $CD_{chiroptic}$ and $LD_{natural}$ are the intrinsic material values and correspond to the 1st-order terms in the equations ($CD$ and $LD` \sin 2\theta - LD \cos 2\theta$, respectively), which are modified by 2nd-order effects ($CD_{LDLB}$ and $LD_{CDLB}$, respectively). For the measured CD intensity, the 2nd-order modification is commonly called "$LDLB$" terms ($LBLD` - LDLB`$). The absorption from these terms has been leveraged to realize NRCD, the preferential absorption of orthogonal circular polarizations through opposite faces of a film (**Fig. 1a**)(*11, 19-23*). Equally, the measured LD intensity has contributions from the intrinsic $LD$ term (or $LD'$, depending on the angle) that is modified by $CDLB$ and $CBLD$ terms (e.g., $CDLB - CBLD$) (*7, 24*). These mixed circular and linear terms ($CDLB$ and $CBLD$) have been previously mentioned in literature, for instance in the works of Schellman and Jensen, but were taken to be negligible under the assumption that the natural LD and LB are orders of magnitude stronger than the CD (*19, 25, 26*). In this work, we revisit these expressions to show that these mixed chiral-linear terms can be leveraged to realize NRLD, the preferential absorption of orthogonal linear polarizations through opposite faces of a film (**Fig. 1b**).

To illustrate how the $LDLB$, $CDLB$, and $CBLD$ terms infiltrate CD and LD spectra, we present idealized materials and their effect on the measured signal. Light incident on an isotropic solution of chiral molecules or aggregates (**Fig. 1c**) will result in a CD signal but no LD signal; even if the individual chiral species possess an anisotropic electric transition dipole moment (TDM), no LD will be measured because the TDMs are randomly oriented. In contrast, consider an anisotropic film composed of linearly aligned oriented chromophores (**Fig. 1d**). Because the TDMs of the chromophores are oriented, an LD signal is measured. And because these are achiral chromophores that are not optically active, there will not be a CD signal measured (*27*).

Next, consider an "LDLB" material (**Fig. 1e**). These materials possess linear anisotropy (LD & LB) in multiple directions that combine to produce a CD signal even in the absence of intrinsic CD. The CD for these materials is polarimetrically nonreciprocal in that the signal inverts upon sample flipping (**Fig. 1e,** front vs. back CD spectrum). This signal inversion is caused by the inversion in the angle between the LD and LB` axes (**Fig. 1e,** schematic). In a Mueller matrix polarimetry (MMP) measurement, the signal inversion can be recognized by the $M_{03}$ and $M_{30}$ matrix elements possessing equal magnitudes but opposite signs (*28*). It's important to note that an LDLB sample *can* and often will have an LD signal, although this isn't explicitly required to observe a NRCD signal (*7, 21*).

Finally, consider a film composed of chromophores that are both optically active and assembled into a highly aligned, linearly anisotropic film (**Fig. 1f**). Under these conditions, it is possible to create a NRLD signal that inverts sign upon sample flipping. In a MMP measurement, this effect would be recognized by the $M_{01}$ and $M_{10}$ matrix elements possessing equal magnitudes but opposite signs. This effect has not previously been appreciated because the strength of this effect is determined by the relative magnitudes of the optical effects, and if the CD and CB are orders of magnitude weaker than the linear effects – which is the typical case – there will be negligible contributions of the NRLD signal to the measured spectra (*19, 25, 26*).

**Results and discussion**
To experimentally realize NRLD, we employed CdS magic-size cluster (MSC) films, as they fill the requirements for possessing both strong chiroptic effects and linear anisotropy (*16, 29-31*). To form optically active and linearly anisotropic films, we process solutions of CdS MSCs through confined evaporation, into banded films with twisted fiber that possess exceptionally large LD and CD from exciton coupling (**Fig. 2a,b**)(*16,*

*30, 31*). The resultant CdS MSC films possess all 6 polarized optical effects (LD, LB, LD`, LB`, CD, and CB) with large magnitudes (CD dissymmetry exceeding 1 and reduced-LD exceeding 0.6) (**Fig. 2a,b**)(*16, 31*).

Measuring a CdS MSC film with the bands oriented at 45° to vertical enables the experimental realization of NRLD (**Fig. 2d**). The film is measured at four orientations, following the methodology of our previous work which isolated $CD_{chiroptic}$ from LDLB contributions (eq. 1b)(*16*). When measured in the forward direction a positive LD signal is recorded (**Fig. 2d**, brown). When the sample is flipped, reversing the illumination direction, the sign of the LD spectra inverts to a negative LD value (**Fig. 2d**, dark blue), marking, to our knowledge, the first-ever experimental realization of NRLD. To confirm that this behavior is not an artifact of the measurement, we rotated the sample 90°; the sign of the forward-illuminated LD inverts from the 0° measurement, as it should for any LD-dominant material, and then we flipped the sample orientation and the sign of the LD inverts upon sample flipping (**Fig. 2d**, orange vs light blue), perfectly mirroring the same sequence of measurements made at the 0° position. Reversing the sense of the wavevector, which is equivalent to rotating the sample, changes the sign of the linear prime effects, LD` and LB`, ultimately inverting the sign of the measured NRLD (**Fig. 2c**)(*28*).

We can isolate the NRLD components measured through the front and back with the following expressions (derivation in SM):

$$LD_{CDLB,Front} = \frac{1}{2}(LD_{measured,Front} - LD_{measured,Back}) \quad (3)$$

$$LD_{CDLB,Back} = \frac{1}{2}(LD_{measured,Back} - LD_{measured,Front}) \quad (4)$$

Where $LD_{CDLB,Front}$ is the contribution of the NRLD term in front illumination, and the $Back$ term is in reverse illumination. Calculations of our spectra using these expressions yield symmetric spectra of opposite sign (**Fig. 2e**, solid lines).

To confirm the origin of this nonreciprocal effect we can directly model the behavior using representative excitonic lineshapes for the dichroic optical effects. The LD` is modeled using a Gaussian at the transition energy, the CD is modelled with two split Gaussians of opposite sign, and their birefringent counterparts, LB` and CB, are computed through a Kramers-Kronig transformation (**Fig. S5**, model details in SM). These lineshapes were chosen to match the excitonic effects of the MSCs: the total absorbance values from the dichroic effects were kept at levels that were less than or equal to the experimental absorbance (**Fig. 2g**), and the magnitudes of the LD` and CD were chosen to match the values recorded for this film (**Fig. S2, S5**)(*31*). Then using the expression for the measured LD, $LD_{measured}$ (eq. 2a), we compute the LD measured by the instrument from the front of the sample (**Fig. 2e**, orange dashed spectrum). Flipping the sign of the linear prime effects causes $LD_{measured}$ to flip sign (**Fig. 2e**, blue dashed spectrum). This model achieves good agreement with the measurements: it replicates not only the lineshape and magnitude of the LD effects, but also the nonreciprocal nature of the LD.

Kramers-Kronig relations connect the real and imaginary parts of analytic complex functions to one another, in the context of polarimetry connecting dichroic and birefringent effects to one another (*32, 33*). Thus, the above experimental result and numerical model reveal yet another nonreciprocal optical effect, *nonreciprocal LB* (NRLB) (**Fig. 2f**). The NRLB presents a bisignate lineshape centered about the absorption maximum, again finding good agreement between experiment and the proposed model. NRLB can be thought of as an inversion between the "fast" and "slow" axes of a material, and the absorption-centered bisignate lineshape is expected, based on a driven oscillator model (*31*).

To test the generality of this NRLD realization, we examined films made from CdSe and CdTe MSCs and found NRLD in these samples as well (**Fig. 3a**) (*31, 34, 35*). These films present chiral and linear anisotropy with similar forms and origins to the CdS MSC films, and the resultant NRLDs have similar lineshapes, closely mimicking the absorption lineshapes. Demonstrating NRLD with three unique semiconducting MSCs highlights the robustness of the effect and its potential to be engineered in a wide range of other systems. As a control, we measured the linear dichroism of several linearly-aligned, optically inactive samples, which showed no NRLD response, as predicted by the Stokes-Mueller expression, eq. (2a) (**Fig. S12, S13**).

Based on equation 2a, the NRLD should present a strong θ orientation-dependence. At intervals of $n\frac{\pi}{2}$, where $n$ is an integer, the NRLD ($LD_{CDLB}$) reaches its maxima and minima, and $LD_{CDLB}$ smoothly varies between these angles, reaching 0 at interval angles of 45° (**Fig. 3b**, left). Because $LD_{CDLB}$ smoothly varies, this angular dependence presents opportunity to turn the NRLD "on" and "off" by rotating the film. At the angles where $LD_{CDLB}$ falls to 0 (e.g., $\theta = \frac{\pi}{4}$), a LD measurement will be dominated by the natural LD within the film. When measured through the back face of a film, the angle dependence is maintained, but the $LD_{CDLB}$ inverts its sign to be opposite sign but equal magnitude to a front measurement (**Fig. 3b**, right).

The lineshape of the $LD_{CDLB}$ term is determined by the lineshape of the LD and CD contributions. All three MSC species studied in this work exhibit a bisignate CD, arising from their exciton-coupled chirality, and a Gaussian LD from their linearly aligned transitions. The combination of these optical effects leads to a monosignate $LD_{CDLB}$ contribution, with a Gaussian-like lineshape centered about the absorption maximum

(experimental results in **Fig. 2e, 3a,** and modeled in **Fig. 3c**). If, however, the natural CD contribution is more classically defined by a monosignate CD, then the model instead predicts that a bisignate $LD_{CDLB}$ lineshape should be realized. This bisignate $LD_{CDLB}$ still inverts upon sample flipping, but it also changes sign on either side of the absorption maximum, which expands the functionality of the NRLD effect. Further exploration of the realizable NRLD lineshapes could improve polarization multiplexing: creating unique polarization states at unique energies.

These nonreciprocal linear effects can occur in linearly-polarized luminescence (LPL) measurements (**Fig. 3d**). Assuming that the luminescent optical axis of a sample is the same as that of its absorption enables us to use Stokes-Mueller calculus to derive an expression that describes the polarization-state of light emitted after unpolarized excitation (details in SM) (*36*). This expression has a similar form to equation (2a), requiring strong chiral and linear anisotropy along the diagonal. Modeling a film with properties matching those of our MSC films predicts the emergence of nonreciprocal LPL, evidenced by asymmetries in the calculated luminescence spectra lineshapes (**Fig. 3d**, dashed spectra). Experimental LPL measurements of a CdS MSC film through opposite faces confirms the existence of nonreciprocal LPL (**Fig. 3d**, solid spectra, **Fig. S16-S20**). The moderate differences between the predicted and experimental spectra likely lie in the lineshapes used for the PL. As a control, measuring an achiral, anisotropic film of CdS nanorods yields strong LPL but no evidence of a nonreciprocal LPL component, as predicted by the derived expressions, substantiating this discovery in the MSCs (**Fig. S21-S23**).

To demonstrate the impact of these $2^{nd}$-order effects on the spectra and assess our ability to model and predict them, we compare experimentally measured samples with complex lineshapes to models that incorporate both $1^{st}$-order LD and NRLD contributions. The NRLD terms, in addition to higher order terms outlined in our previous work, cause significant distortions to spectra, influencing both their intensity and lineshape (derived in SM)(*37*). We model the output LD signal, $LD_{measured}$, and compare this model to an experimental spectrum, estimating the sign, magnitude, and lineshape of the optical effects based on the measurement.

Consider first a MMP measurement of a CdS MSC film (**Fig. 4a**). The $M_{01}$ matrix element, which is reported as LD on standard spectrometers, has a doublet lineshape (**Fig. 4a**, left). Through analytic inversion, this lineshape is converted to a Gaussian centered about the absorption maximum (**Fig. 4a**, right)(*38*). As mentioned above, a sample with purely NRLD contributions will result in $M_{01}$ and $M_{10}$ matrix elements with equal magnitudes but opposite signs, which we have observed in our MMP data (**Fig. S9, S10**).

Now consider two characteristic "extreme" lineshapes found in experimental LD spectra ($LD_{measured}$). The first spectrum has a "doublet" feature, two broad peaks centered about the absorption maximum, which is observed in MSC films made of CdS, CdSe, or CdTe (**Fig. 4b**, solid spectrum). The second spectrum has a lineshape composed of three peaks. There is a positive peak centered at the absorption maximum, and on either side of this peak there are two negative peaks (**Fig. 4b**, solid spectrum). Using the models we developed, we were able to closely replicate each of these lineshapes (**Fig. 4b,c**, dashed spectra)(**Fig. S25**).

From fitting the spectra with the models, we are able to extract the material's intrinsic LD. To obtain this value, we use the 1$^{st}$-order expansion for $LD_{measured}$ (**Table SXX**): this expression relates LD and LD` to the angle the film is rotated to compute the LD the instrument *should* measure. From this methodology, we find that the LD extracted from each of the models is a simple Lorentzian lineshape, centered about the absorption maximum (**Fig 4b,c**, right), finding close agreement with the analytical inversion of MMP data (**Fig. 4a**, right). The ability of the models to accurately replicate the experimental LD lineshapes further supports the argument that nonreciprocal and higher-order effects influence LD measurements in state-of-the-art samples.

**Wave propagation in homogeneous nonreciprocal media**
Now, we establish a conceptual picture of how these complex light-matter interactions mix to influence the output signal by examining their matrix descriptions (**Fig. 5**). Consider first a measurement of CD for an isotropic, optically active sample (**Fig. 5a**). Right-circularly polarized (RCP) and left-circularly polarized (LCP) light are incident upon the sample, and, because the sample possesses CD, the RCP and LCP light are absorbed to different degrees. Because the sample also possesses CB, the RCP and LCP light will transmit through the sample at different speeds. The light will still emerge from the sample as circularly polarized. The net result of this experiment is the measurement of a CD signal.

Next consider the measurement of CD in an anisotropic sample exhibiting NRCD through LDLB effects (**Fig. 5b**). Conceptually, these samples have been explained by imagining RCP and LCP light passing through an LB material followed by an LD` material (**Fig. S27a**)(*22*). In this framework, the LB material converts the circularly polarized light into elliptically or diagonally polarized light, and the LD material then absorbs one polarization more strongly, resulting in preferential absorption of RCP or LCP and, in turn, producing a CD signal. However, this description is inaccurate. It fails to account for nonreciprocal effects: running the experiment in the reverse direction would not generate a CD signal. Moreover, it oversimplifies the behavior of LDLB materials, where dichroic and birefringent effects occur simultaneously and repeatedly throughout the medium, breaking the homogeneous medium assumption used to derive the expression.

An LDLB material is distinctly different in this manner than an LD and an LB material stacked together (**Fig. S27a**). In a CD measurement of an LDLB material, the incident LCP and RCP waves are quickly transformed into elliptically polarized light oriented differently and with different intensities, ultimately yielding a differential absorption of the orthogonal circular polarizations and a CD signal (**Fig. 5b**)(*28*). The exact manner of how the light is transformed depends on the geometry and magnitude of the dichroic and birefringent effects. This picture also correctly depicts the nonreciprocal behavior of an LDLB material: if the medium is flipped with respect to the incident beam, the prime effects change sign, which will in turn change the sign of the CD signal the instrument measures (*28*).

Shifting our attention now to LD measurements, we consider first a linearly anisotropic sample, possessing only LD and LB along one axis (**Fig. 5c**). Vertically and horizontally polarized light are both incident upon the sample, and depending on the orientation of the film, one is absorbed and retarded to a greater degree by the dichroic and birefringent effects, respectively. While the phase and intensity of each of the incident polarization states will be altered, the light will still emerge linearly polarized. The net result of this measurement is an LD signal.

Lastly, we consider now a NRLD sample that possesses linear prime effects, LD` and LB`, and circular effects, CD and CB (**Fig. 5d**). This sample possesses no horizontal or vertical anisotropy, so naturally an LD signal is not expected to be measured based on the symmetries of the material and the differential Mueller matrix. However, this sample will surprisingly yield an LD signal, as demonstrated herein. The incident linearly polarized light beams (vertical and horizontal) are rotated, retarded, and absorbed simultaneously by the birefringent and dichroic effects characteristic of each infinitesimally thin differential Mueller matrix, ultimately resulting in the emergence of elliptically polarized light. There will be differential absorption of the incident polarization states, yielding an LD signal.

There are several parallels we can draw between the LD artifacts derived in this work and the well-established LDLB effects that yield NRCD spectra. One differentiator between these effects and their "true" or "natural" counterparts (optical activity vs. LDLB and linear anisotropy vs. CBLD`/CDLB`; alternatively, $1^{st}$-order vs $2^{nd}$-order) is that in the "true" case, the light emerges with its polarization state preserved (*28*). The other differentiator between the reciprocal and nonreciprocal effects is polarimetric reciprocity:

the "true" optical effects are constant upon wavevector reversal, whereas the nonreciprocal effects invert upon reversing the sense of the wavevector.

Finally, we propose design rules, identifying crystal classes that can display NRLD (**Fig. 5e**). There are 15 crystal classes that possess optical activity, and 19 that possess uniaxial linear anisotropy (*39*). The overlap between these two sects of crystal classes reveals 8 crystal classes that should display NRLD (**Fig. 5e**). Realizing NRLD within these systems requires, however, a careful measurement along the correct optical axis at the correct orientation, as NRLD possesses an orientation dependence (**Fig. 3b**). We hope that this parameter space will motivate researchers to more carefully re-investigate century-old materials such as quartz that should present NRLD (*40*).

**Conclusions and outlook**
In summary, we have used Stokes-Mueller formalism to derive expressions for the LD that is measured on a standard CD spectrometer. The expressions we derived possess, in addition to the LD terms that one hopes to measure in an LD experiment, nonreciprocal terms caused by chiral-linear interference effects. We then verified the existence of the nonreciprocal effect in real LD experimental measurements using CdS MSC films, measuring NRLD for the first time. Modelling these NRLD contributions achieved exceptional agreement between experiment and theory. We substantiated this finding by isolating NRLD contributions in experimental measurements of CdSe and CdTe, proving the NRLD phenomenon is general. We presented design rules for the effect, how it can be realized in other materials and its dependence on orientation and lineshape of the optical effects. Finally, we used these findings to predict nonreciprocal LPL, and realized this theoretical prediction experimentally in CdS MSC films. We discovered an overlooked effect, NRLD, that occurs in ordinary materials that possess chiral and linear anisotropy.

Looking forward, we hypothesize that NRLD can be realized in a wide variety of self-assembled materials through careful design. To this end, the NRLD effect presents a unique opportunity: it can be realized at any wavelength, if the optical effects occur at that wavelength, in a passive material. Discovering the NRLD effect in CdS films at UV wavelengths speaks to this, as realizing UV metamaterials is difficult. Additionally, the NRLD achievement within this work occurs within a single material, greatly simplifying sample preparation. It is possible too that there are separate coherent and incoherent contributions to the NRLD effect, as has recently been observed in NRCD films, making this another area for exploration (*41, 42*).  Our results open previously unexplored possibilities for bottom-up design of compact, scalable nonreciprocal films in familiar materials through established techniques.

**On the term 'Nonreciprocity'**
The nonreciprocal concept being described throughout this manuscript is in regards to *polarimetric reciprocity*, rather than *electromagnetic reciprocity* (*i.e.*, *Lorentz reciprocity*) (*7, 43*). Polarimetric reciprocity can easily be broken, whereas electromagnetic reciprocity cannot (*43*).

**Acknowledgements:**
We thank Dr. Mark August Pfeifer for lending us his rotatable Glan-Taylor polarizer, Dr. Yuan Yao for help developing the Stokes-Mueller calculus used in the derivation, and the Diamond Light Source B23 team for facilitating MMP measurements.
**Funding:** This work was supported in part by the National Science Foundation (NSF) under award nos. CMMI-2120947, DMR-2344586, and CHE-2003586. O.A. acknowledges support by PID2022-138699OB-I00(MCIU/AEI/FEDER, U) funded by the Spanish Ministry of Science, Innovation and Universities (MCIU) and co-funded by the European Regional Development Fund (ERDF, EU). This work made use of the Cornell Center for Materials Research shared instrumentation facility. MMP data was collected at the Diamond Light Source B23 beamline under proposals SM32994 and SM34669.
**Author contributions:** T.J.U. and R.D.R. conceived and planned the study. R.D.R. directed the project. T.J.U. derived the expressions for CD and LD measurements with assistance from D.J.G., O.A., and R.D.R. T.J.U. and D.J.G. synthesized and characterized MSCs and MSC films. T.J.U. modelled the optical response of NRLD films with input from D.J.G. and R.D.R. All authors developed the conceptual model for the nonreciprocal effects. T.J.U. and R.D.R. authored the paper with contributions from all other authors.
**Competing interests:** The authors declare no competing financial interests.
**Data and materials availability:** All data are available in the manuscript or in the supplementary materials.

**Fig. 1:**

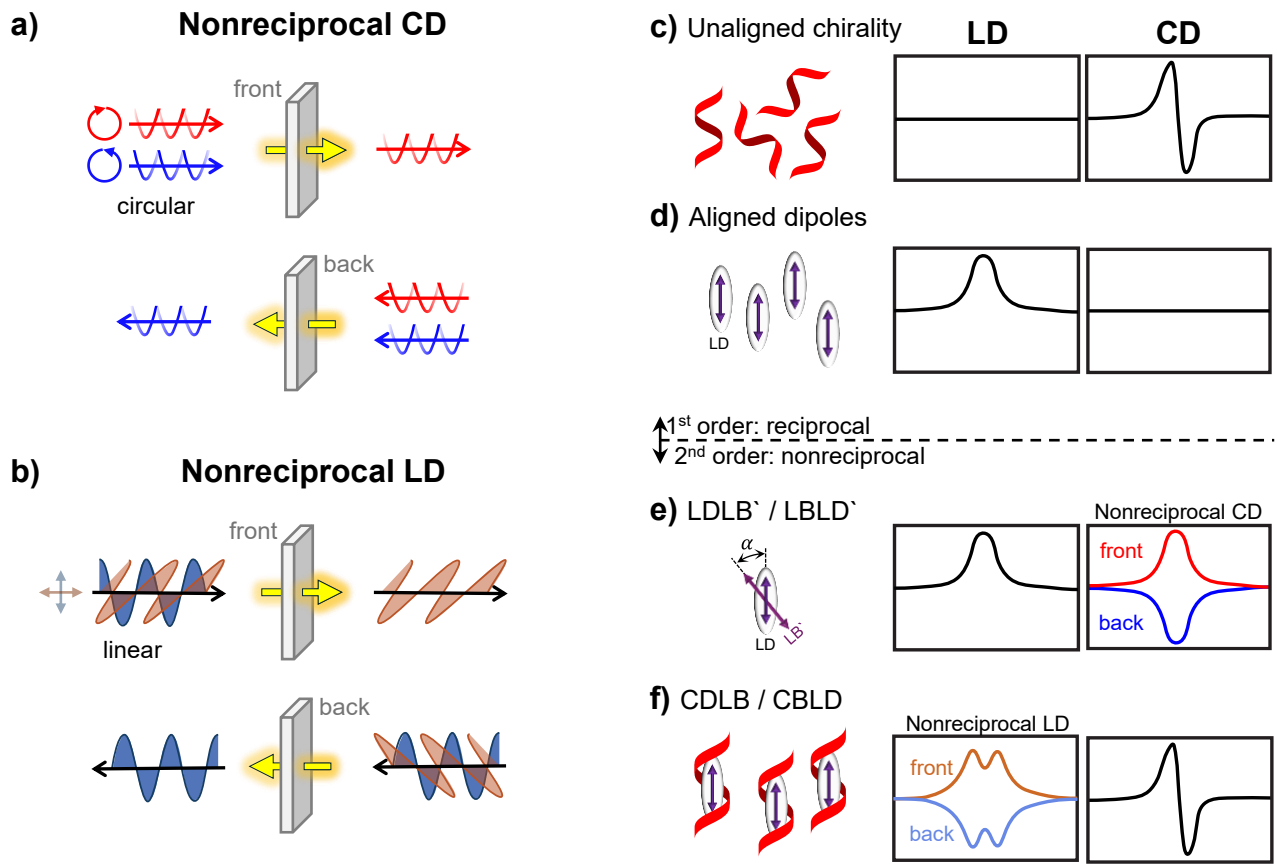

**Figure 1.** Nonreciprocal circular and linear dichroism. **a-b)** Schematic depicting nonreciprocal CD and LD: front side illumination results in polarization states that are orthogonal to light emanating from back side illumination. **c)** Measurements of an isotropic solution of chiral molecules yields a zero LD and a nonzero CD, with a lineshape depending on the origin of the optical activity. **d)** Measurements of a linearly anisotropic sample yields nonzero LD and zero CD. **e)** Measurements of an LDLB sample can yield zero or nonzero LD, depending on the nature of the sample, but will present a non-zero nonreciprocal CD signal that inverts upon sample flipping. If the sample is optically active, the LDLB effects will distort the CD spectrum with nonreciprocal effects. **f)** Measurements of a chiral and linearly anisotropic sample should present LD and CD spectra characteristic of the linear anisotropy and optical activity. And, the interactions result in novel, nonreciprocal LD signals that invert upon sample flipping. The interference effects can also distort the spectra, as shown here.

**Fig. 2:**

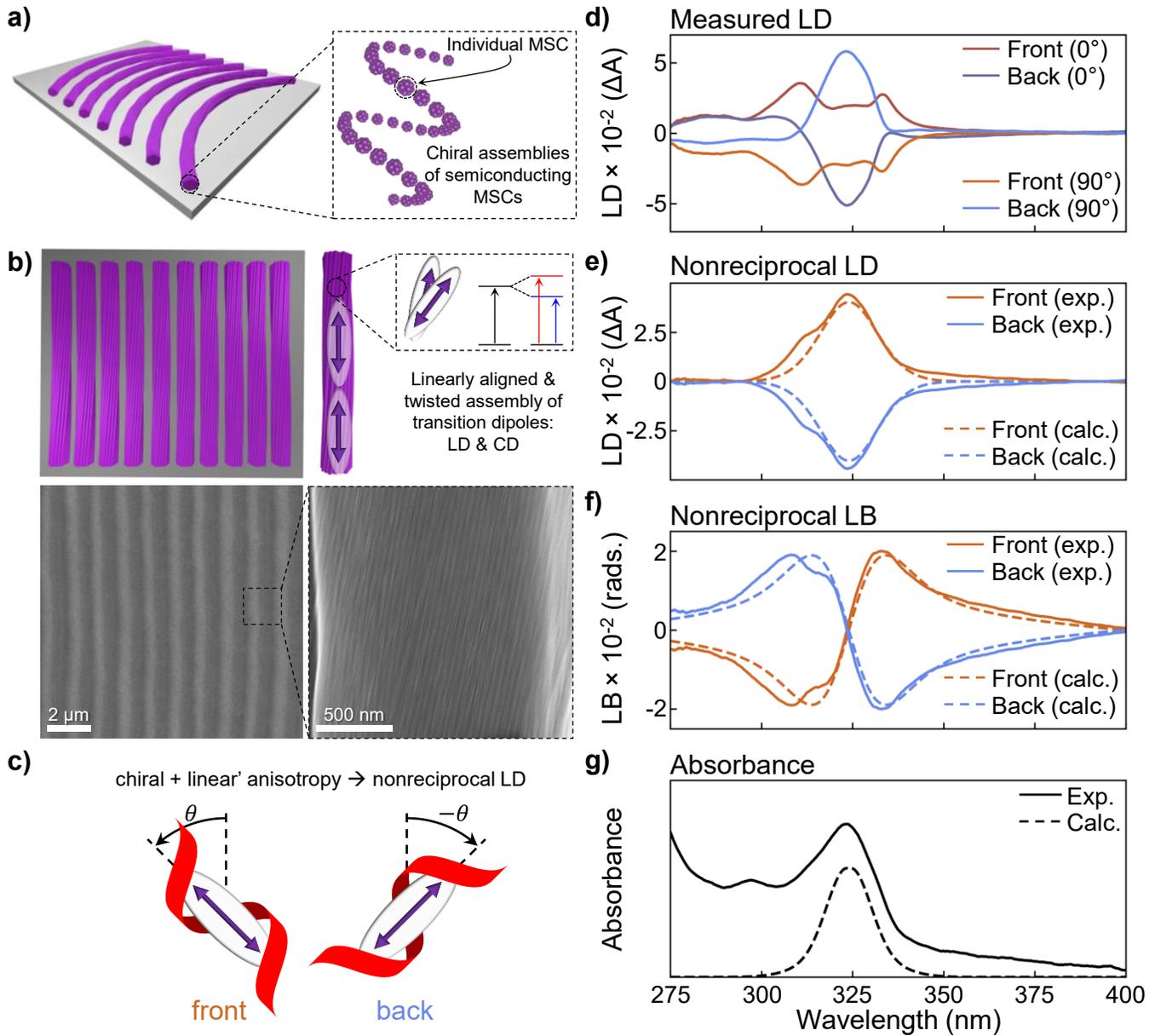

**Figure 2.** Nonreciprocal linear dichroism in CdS MSC Films. **a)** Achiral CdS MSCs form assemblies in solution and, through controlled evaporation, form macroscopic banded films that possess strong LD and CD. **b)** Scanning electron micrographs and schematics of an MSC film. Macroscopically, aggregates of MSCs are assembled and helically twisted, side-by-side on a substrate. Within these twisted aggregates are hexagonally arranged, preferentially oriented MSCs, resulting both linear (LD, LB, LD`, LB`) and chiral (CD, CB) optical effects. **c)** Schematic depicting the origin of nonreciprocal linear dichroism and birefringence. The nonreciprocal effect originates from a sign inversion of the prime effects upon sample flipping. Because the linear prime effects (LD` and LB`) are defined with respect to the experiment geometry, the sign of these effects invert upon sample flipping. The circular effects (CD and CB) are rotationally invariant, so their sign is independent of sample flipping. **d)** Measured LD values from sample at 0° and 90° when illuminated on the front side and back side. For a fixed angle, the sign of the LD inverts from the front to back. **e)** Experimental isolation and numerical modeling of nonreciprocal LD terms. Using equations (3) and (4), the contributions of the nonreciprocal LD from the front and the back are isolated from the individual scans. For the numerical modeling of nonreciprocal LD terms, the sample is assumed to have moderate linear prime and circular effects. To model sample flipping, the sign of the linear prime effects are inverted. **f)** Numerically modeled nonreciprocal LB, using the same parameters used to match experiment in panel **e**. **g)** Experimental and numerically modeled absorbance.

**Fig. 3:**

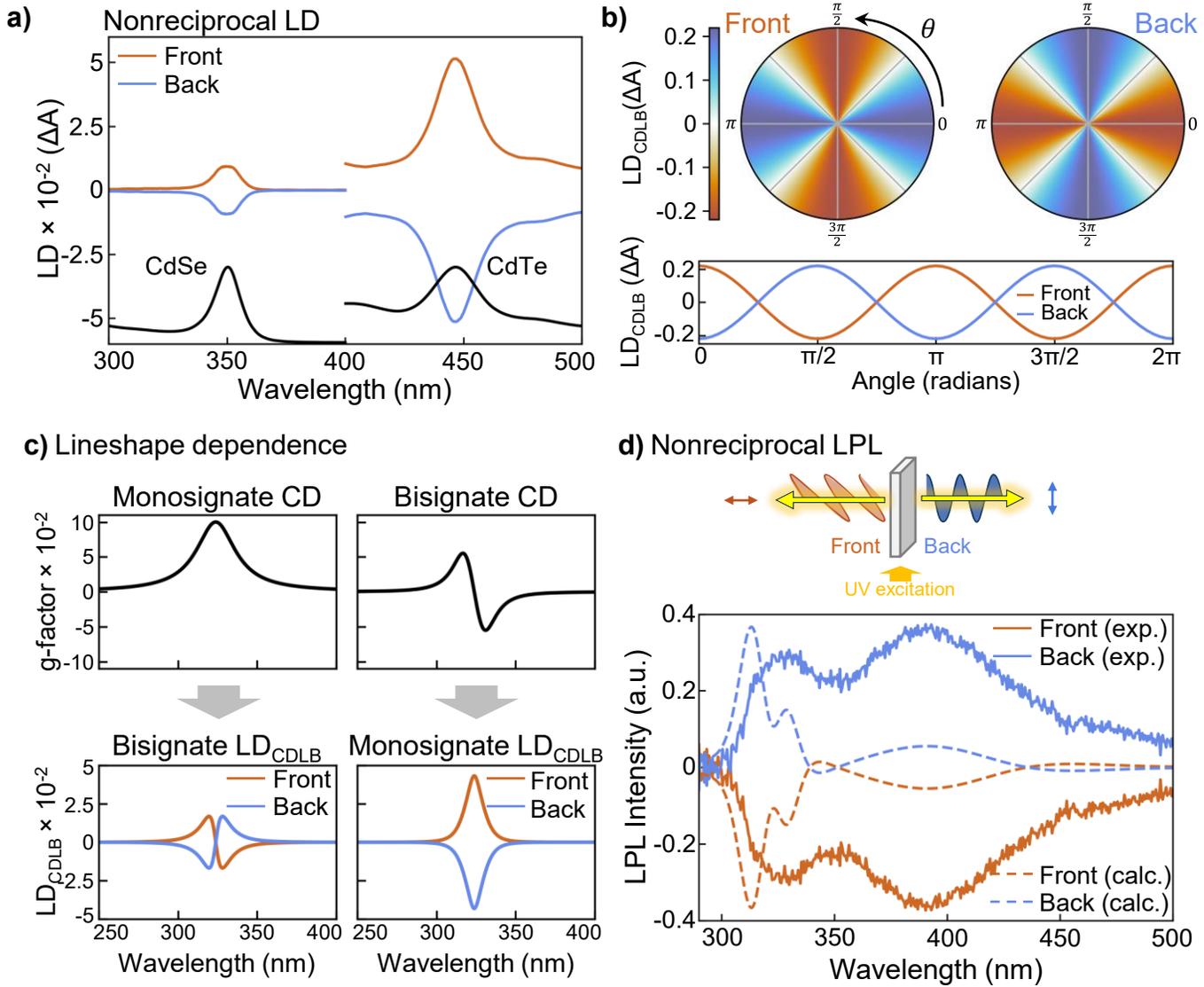

**Figure 3.** Extended functionality of nonreciprocal linear dichroism. **a)** Experimentally-isolated nonreciprocal LD from films composed of CdSe and CdTe MSCs, and the measured absorption (at bottom). **b)** Numerically derived rotational-dependence of the nonreciprocal LD terms. As sample is rotated the $LD_{CDLB}$ term follows a sinusoidal pattern, and backside illumination has inverted behavior compared to the frontside illumination. **c)** Lineshape dependence of the nonreciprocal LD effect on the CD profile. Assuming a Gaussian LD and a Monosignate CD yields a bisignate nonreciprocal LD absorption profile. But if the LD is gaussian and CD is bisignate then a gaussian nonreciprocal LD profile emerges. **d)** Nonreciprocal emission from the front side and backside of a CdS MSC film. Emission profiles are linearly inverted for the two sides. The sample was excited with a 272.5 nm laser source (5 nm bandwidth), and the spectra presented are the difference between the front and back LPL after normalizing by the unpolarized excitation.

**Fig. 4:**

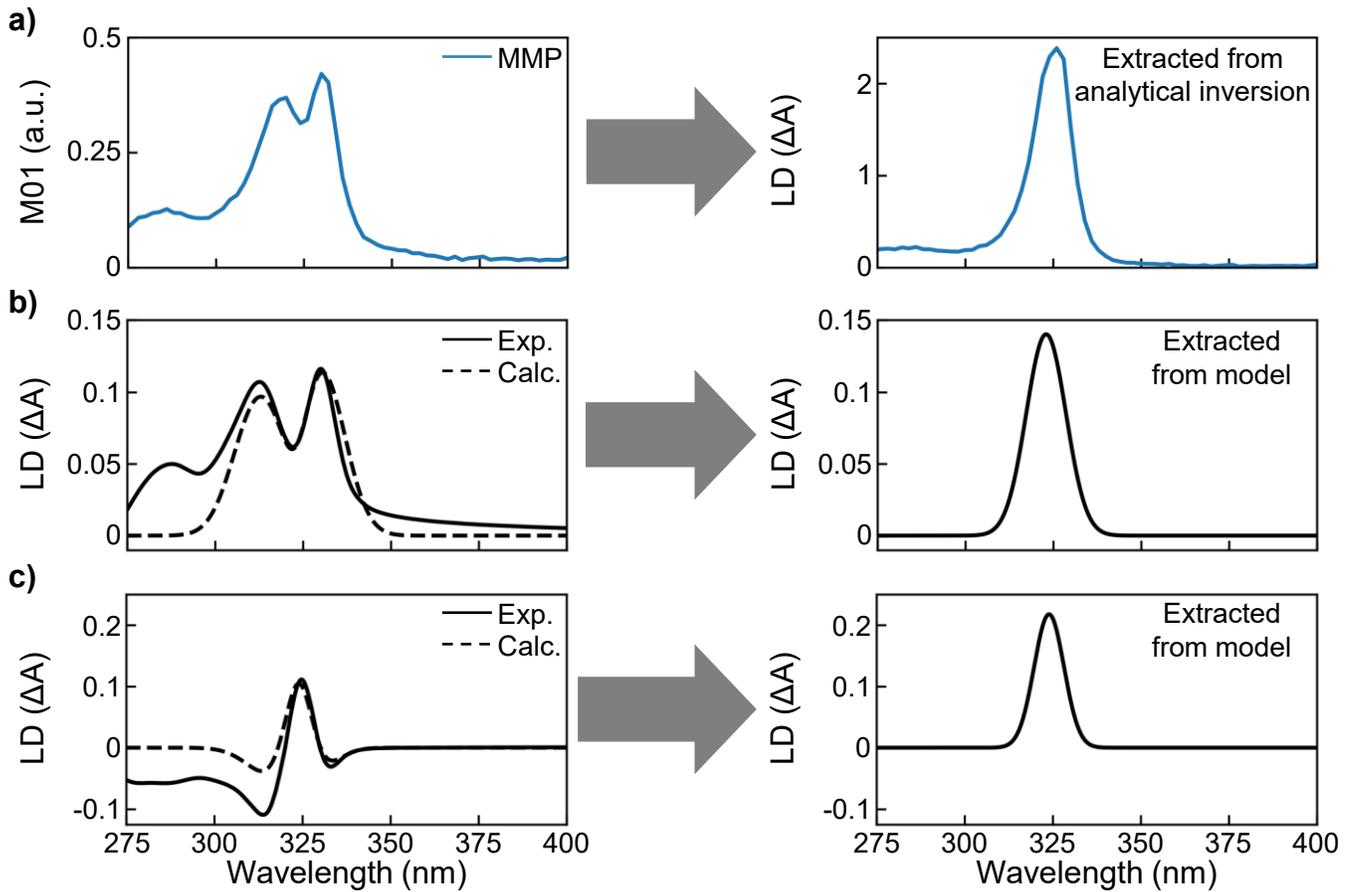

**Figure 4.** Examples of showing methods to model the experimental profiles of LD from CdS MSC samples. The experimentally-measured LD display artifacts that can be accounted for with simple modeling. **a)** Left: M01 matrix element collected from a CdS MSC film through Mueller matrix polarimetry (MMP). Right: The LD extracted from this sample using the analytical inversion method, which corrects for higher-order contributions to the spectrum. This spectrum represent the natural LD contribution. **b)** Left: experimental and calculated LD spectrum of an MSC film presenting LD artifacts with a "doublet" lineshape. Right: The "corrected" LD of this MSC film, extracted from the modelled spectrum at left. **c)** Left: experimental and calculated LD spectrum of an MSC film presenting LD artifacts with a more complex lineshape. Right: The "corrected" LD of this MSC film, extracted from the modelled spectrum at left.

**Fig. 5:**

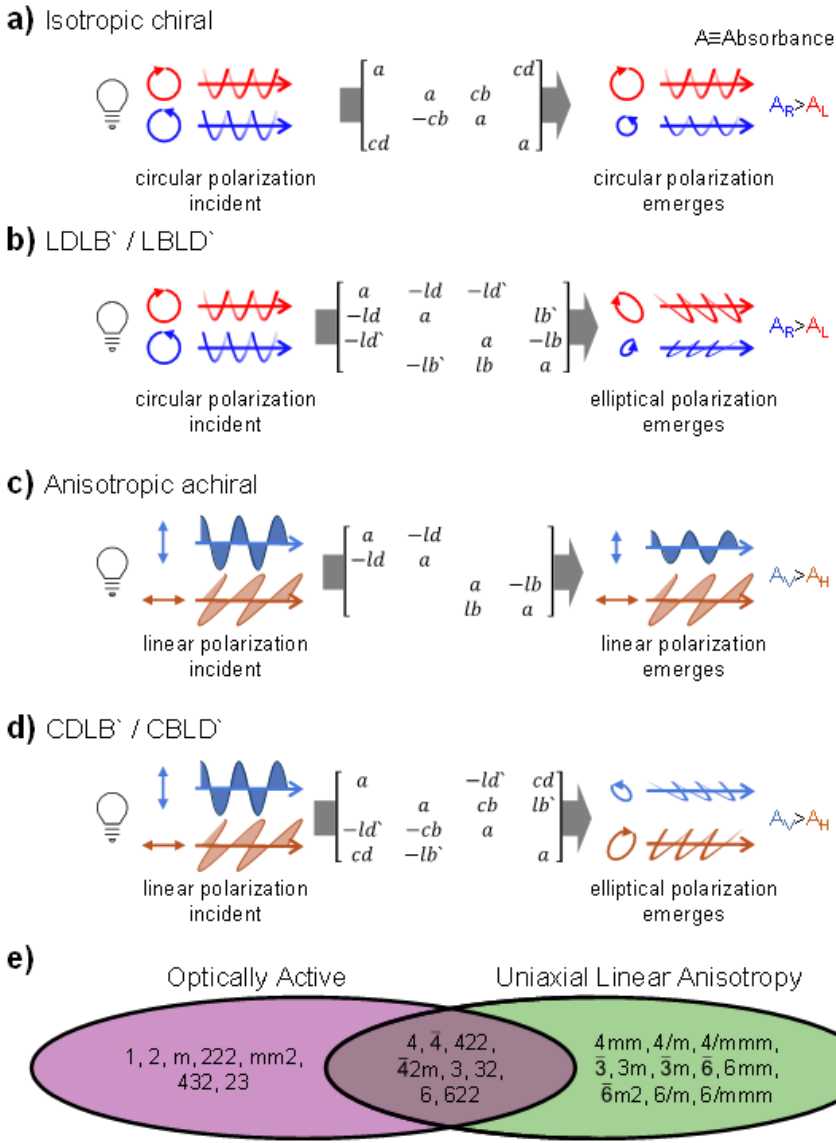

**Figure 5.** Differential Mueller matrix formalism reveals the difference between optical activity and LDLB effects in a CD measurement and linear anisotropy and CBLD effects in an LD measurement. **a)** Schematic depicting the circular dichroism of an optically active, isotropic sample. One of the incident circularly-polarized waves is absorbed to a greater degree, and the polarization state is preserved as circularly-polarized. **b)** Schematic depicting the circular dichroism of an anisotropic sample with LDLB effects but no chirality or optical activity. Much like above, one of the incident circularly-polarized waves is absorbed to a greater degree due to LDLB interactions, yielding a CD signal. However, the polarization state is no longer preserved as circularly-polarized because of the dichroic and birefringent effects, instead being projected to become elliptically- or linearly-polarized. **c)** Schematic depicting the linear dichroism of a linearly anisotropic sample. One of the incident linearly-polarized waves is absorbed to a greater degree, and the polarization state is largely preserved as circularly-polarized. **d)** Schematic depicting the linear dichroism of an anisotropic sample with optical activity and linear effects oriented along the diagonal. Despite possessing no horizontal or vertical anisotropy, one of the incident linearly-polarized waves is absorbed to a greater degree due to CBLD` and CDLB` effects, yielding an LD signal. However, the polarization state is no longer preserved as vertically- or horizontally-polarized, instead being rotated by the optical activity and absorbed to varying degrees by the diagonal anisotropy. **e)** Crystal classes that display optical activity and uniaxial linear anisotropy, the overlap of which should result in NRLD and nonreciprocal linear emission.

# Supplemental Materials

**Title:** Realizing Nonreciprocal Linear Dichroism and Emission from Simple Media

**Authors:** Thomas J. Ugras[1,2], Daniel J. Gracias[3], Oriol Arteaga[4], Richard D. Robinson[2,3]*

**Affiliations:**

[1] School of Applied and Engineering Physics, Cornell University, Ithaca, NY, USA.

[2] Kavli Institute at Cornell for Nanoscale Science, Ithaca, NY, USA.

[3] Department of Materials Science and Engineering, Cornell University, Ithaca, NY, USA.

[4] Departament de Física Aplicada, Universitat de Barcelona, IN2UB, Barcelona, Spain.

*corresponding author: rdr82@cornell.edu

## Materials and Methods:

### Chemicals / Materials

Cadmium oxide (Sigma Aldrich, 99.5%, trace metal basis), oleic acid (Thermo Scientific, 90%, tech. grade), oleylamine (Sigma Aldrich, 70%, tech. grade), 1-octadecene (Thermo Scientific, 90%, tech. grade), sulfur powder (Sigma Aldrich, >99%), selenium powder (Alfa Aesar, 99.999% metal basis, -200 mesh), tellurium powder (Sigma Aldrich, 99.997% metal basis, -30 mesh), tri-octyl phosphine (Sigma Aldrich, 97%), tri-butyl phosphine (Oakwood Chemical, 93.5%), tri-octyl phosphine oxide (Aldrich, 99%), n-octadecylphosphonic acid (PCI Synthesis, 98%), hexanes (VWR, >97%, ACS grade), acetone (Macron Fine Chemicals, 99.5%, ACS grade), toluene (VWR, >97% ACS grade), toluene (Thermo Scientific, 99.8% anhydrous), chloroform w/ amylene (Beantown Chemical, 99.8%), methanol (Fisher Scientific, 99.8%, ACS grade), copper(II) acetylacetonate (Aldrich, ≥99.99% metal basis), dibenxyl ether (TCL, >95%), 1-dodecanethiol (Aldrich Chemica, ≥98%), 1-octadecanethiol (Sigma Aldrich, 98%), glass microscope slides (VWR, 75x25x1 mm), Glan-Taylor polarizer (Harrick PGT-S1V)

### CdS MSC synthesis

Cadmium sulfide (CdS) nanoclusters were synthesized using previously published protocols (*16, 29-31, 44*). Cadmium oxide (10 mmol, 1.28 g) and oleic acid (32 mmol, 10 mL) are mixed in a 50 mL three-necked round bottom flask and attached to a Schlenk line. This mixture is constantly stirred using a magnetic stir bar and stir plate, then degassed with vacuum at 50 C until no bubbles remain, up to an hour, before being put under N2. The mixture is then heated to 160 C and allowed to react for at least 1 hour. The reaction product is allowed to cool to 105 C and is again degassed with vacuum for an hour. Then, the reaction product is put under N2 and allowed to cool to 60 C.

Elemental sulfur (5 mmol, 0.16 g) and a magnetic stir bar are added to a scintillation vial and loaded into a N2 glovebox. Tri-octyl phosphine (4.5 mmol, 2 mL) is added to the sulfur using a syringe bearing a hypodermic needle and stirred on a magnetic stir and hot plate at ~80 C for about 10 minutes, until the elemental sulfur is fully reacted. Once fully reacted, the scintillation vial is removed from the glovebox.

Once the reaction products within the three-neck flask are cooled to 60 C, the tri-octyl phosphine sulfur (TOPS) solution is injected, and the reaction mixture is heated to 140 C and allowed to react for one hour. After reacting, the products are cooled to 60 C before being distributed evenly into two centrifuge tubes. Within the centrifuge tubes, about 15 mL of hexanes and 15 mL of acetone are added to each, leading to the formation of a white precipitate. The vials are centrifuged at 4400 r.p.m. for 5 minutes and the supernatant is discarded before resuspending the precipitate in 15 mL of hexanes again. Once resuspended, the precipitation procedure is repeated by quenching with 15 mL of acetone before being centrifuged again at 4400 r.p.m. for 5 minutes. This procedure is repeated two or more times, until the discarded supernatant is clear and colorless. The precipitate is dried under vacuum for 16 hours, at which point it can be used for self-assembly.

**CdSe MSC Synthesis**

The CdSe fibrils were prepared according to a literature procedure (*34*). A stock solution of tributylphosphine selenide (TBPSe) was prepared by adding selenium powder (19.3 mmol,, 0.63 g) and 1-octadecene (54 mmol, 17.3 mL) to a 50 mL round bottom flask. The mixture was heated to 60 °C and vacuumed until bubbles ceased to evolve. The mixture was then heated to 150 °C and TBP (9.3 mmol, 2.3 mL) was injected into the mixture. The reaction proceeded for 2 h and was then cooled to room temperature and stored in a N$_2$ atmosphere glovebox.

Cadmium oxide (2mmol, 0.254 g), trioctylphosphine oxide (7.6 mmol, 2.93 g), and 1-octadecyl phosphonic acid (3.2 mmol, 1.070 g) were added to a 50 mL three neck round bottom flask on a standard Schlenk line. The ligands were melted at 60 °C and the temperature was raised to 130 °C then vacuumed for 1 h. The reaction flask was then flushed with dry N$_2$ and the temperature was raised to 330 °C for 30 min. The reaction flask was then cooled to 100 °C and vacuumed again for 1 h. The temperature was then raised to 340 °C and 2.32 mL of the TBPSe solution was rapidly injected into the reaction flask. The reaction continued for 2 minutes after injected before it was quenched by adding 5 mL of anhydrous tolened in rapid 1 mL bursts. The heating mantle was removed, and the red opaque product was cooled to 75 °C, transferred to a centrifuge tub, and then centrifuged (4500 rpm, 5 min). The supernatant was discarded, and the precipitate was rinsed with hexane (2 mL). The product was then dissolved in hexane (20 mL) and heated to 50 °C in a hot water bath until the solution became clear. The solution was then centrifuged again (9000 rpm, 5 min), and the supernatant was decanted and retained. Flowing $N_2$ gas was run over the solution until it became a gel. This was then used to fabricate films using the procedure below.

### CdTe MSC Synthesis

The CdTe MSCs were prepared according to a literature procedure with minor additions (*35*). A stock solution of cadmium oleate ($Cd(OA)_2$) was prepared by mixing cadmium oxide (0.77 g, 6 mmol), oleic acid (4.75 mL, 13.5 mmol), and 1-octadecene (5.25 mL), then heating to 100°C under vacuum for 30 mins. Then, the reaction was brought to 250 °C until the solution appeared a clear yellow. After cooling, the solution was stirred at 70 °C in air to prevent solidification. A tri-octyl phosphine telluride (TOPTe) stock solution was prepared by mixing tellurium powder (0.64 g, 5 mmol) and TOP (5 mL, 10.6 mmol) then heating at 100 °C for 30 mins in a $N_2$ atmosphere glovebox. The solution was cooled to room temperature prior to use. The Cd(OA)2 stock solution (2 mL) was mixed with the TOPTe stock solution (1 mL) and diluted in a combination of oleylamine (8.95 mL, 18.7 mmol), TOP (5 mL, 10.6 mmol), and toluene (9.7 mL). The reaction flask was then purged with N2 and the temperature was raised to 100 °C. The reaction continued for 24 hours before being cooled to room temperature. The raw product was then placed in two 50 mL centrifuge tubes and cooled in a refrigerator for 36 hours. A separation layer formed with a clear yellow layer on top and a turbid dark orange layer on the bottom. The mixture was then centrifuged (7500 r.p.m, 3 min) and the supernatant was discarded and the precipitate retained. The precipitate was then redissolved in toluene (3-5 mL) and used to fabricate thin films using the procedure outlined below.

### Cu MSC Synthesis

Cu MSCs were prepared according to a literature procedure with modifications (*45*). Copper(II) acetylacetonate (0.11 mmol, 30 mg) and anhydrous toluene (47.06 mmol, 5mL) were added to a 25 mL three-neck round bottom flask in air. The mixture was then heated to 80 °C. The reducing agent and ligand, 1-dodecanethiol (4 mmol, 1 mL), was injected at this temperature, and the reaction proceeds for 2 min. The reaction was then immediately quenched by placing the round bottom flask into a room-temperature water bath until it cooled to room temperature. The product was then transferred into a centrifuge tube and precipitated with acetone (25 mL) and centrifuged (5000 rpm, 3min). The supernatant was discarded, and the precipitate was redissolved in hexane (5 mL). This gel was then used to fabricate films using the procedure outlined below.

### CdS Nanorod Synthesis

The CdS nanorods used as a control were synthesized following a literature protocol with minor adjustments (*46*). Tri-octyl phosphine oxide (7.1 mmol, 2.73 g), octadecylphosphonic acid (3.19 mmol, 1.07 g), and CdO (1.60 mmol, 0.205 g) were added to a three-neck round bottom glask. The mixture is heated and degassed at 120 °C under vacuum for 30 min before it was heated to 320 °C under $N_2$ to dissolve the CdO. The temperature was then lowered to 180 °C and the solution was vacuumed for 90 min. Separately, a solution of sulfur (1.5 mmol, 0.048 g) in tri-octyl phosphine (1.60 mmol, 0.72 mL) at 8.65 wt% was prepared, with toluene (3.31 mmol, 0.35 mL) used as a solvent. The temperature of the round bottom flask is then lowered to 133 °C, and DI water (7.60 mmol, 0.137 g) was injected. The temperature was then increased to 320 °C, and the S:TOP is injected. The nanorods are grown at this temperature for 90 min. Then, the reaction product was centrifuged (4500 rpm, 5 min), only adding methanol (20 mL) as an anti-solvent. The supernatant was discarded, and a yellow precipitate is kept and dried in a vacuum desiccator. This solid can be suspended in hexanes or toluene and prepared into films. For this work, nanorods were dropcasted into films used.

**MSC Film Assembly**

The various MSCs were self-assembled into hierarchical films using previously published protocols (*16, 30, 31*). Briefly, the MSCs were suspended in a nonpolar solvent (typically chloroform, but hexanes can also be used) at high concentrations, ~20 mg/mL, and allowed to stir under a N2 atmosphere for at least two days. After stirring, the solution will be viscous and gel-like.

Once the nanocluster solutions are prepared, glass slides are prepared with three strips of double-sided Scotch tape in a "Π" shape. Then, a 10-20 µL droplet of nanocluster solution is dropped onto the glass slide (droplet volume is varied depending on desired film size), and a second glass slide is placed atop the first glass slide and gently pushed down to create a uniform droplet and seal, confining the evaporation to one direction, and wetting the surface of each glass slide with the nanocluster solution. Then, the deposited nanocluster solution is allowed to evaporate and precipitate, respectively, leaving behind two films, one on each glass slide.

**Linear and Circular Dichroism Spectroscopy**

*Circular and Linear Dichroism from CD Spectrometer* – Circular dichroism (CD) spectra were collected on a Jasco J-1500 CD spectrometer. The data pitch was selected as 0.2 nm, and the incident light typically ranged from 400 – 250 nm. The digital integration time (D.I.T.) was typically set to 2 seconds for a 100 nm/min scan speed. The CD, linear dichroism (LD), absorbance (A), and high tension (HT) voltage were all collected. If the HT exceeded ~700 V, the spectrum was discarded. Different aperture sizes were used, typically 1, 2, or 4 mm in diameter. A background spectrum was first collected on a bare substrate using the same aperture.

The circular dichroism spectra and values reported within the manuscript were all collected using previously published methods to remove linear anisotropies and inhomogeneities due to the instrument (*16*). Four spectra were measured; two through the front of the sample and two through the back of the sample, each rotated 90 azimuthally relative to one another. These four spectra are then averaged to provide the "true" CD of the sample.

*Linear Dichroism from UV-Vis Spectrometer* – LD spectra were collected using an OceanOptics UV-Vis DH2000 BAL spectrometer equipped with halogen and deuterium bulbs. A rotatable Glan-Taylor polarizer was inserted into the beampath before the sample. Then, five absorbance spectra were collected and averaged together with the polarizer in the horizontal and vertical positions. The difference between horizontal and vertical polarizations was taken and reported as the LD. A background of the glass slide was also used.

## Photoluminescence Spectroscopy

Photoluminescence spectra were collected using an Edinburgh FLS1000 spectrometer. The samples were mounted on the instrument's thin film stage, and the sample's position was adjusted to maximize photon counts. The samples were excited with a Xenon arc lamp set to 272.5 nm with 5 nm bandwidth and no polarizer inserted. Emission spectra were collected from 290 nm to 500 nm with a 0.1 s dwell time and 0.5 data pitch. Spectra were collected for both the front and back of the samples, with the detection side in three conditions: no polarizer inserted, polarizer inserted and oriented horizontally, and polarizer inserted and oriented vertically. Because the counts for a bare, glass side in these conditions were negligible (<500 counts for bare substrates, compared to >>10000 counts for any of the samples studied in this work), no background was subtracted.

The linearly polarized luminescence (LPL) was computed as

$$LPL = \frac{I_V - I_H}{I_{unpolarized}}$$

where $I_V$ is the intensity of vertically polarized light emitted from the sample, $I_H$ is the intensity of horizontally polarized light emitted from the sample, and $I_{unpolarized}$ is the total intensity emitted from the sample (**Fig. 3d, Fig. S16-S23**).

This was done for both faces of the sample (so, in total, a sample is measured in six different conditions). Then, the nonreciprocal LPL was computed as the difference between the front and the back and vice versa.

For the LPL measurements of the CdS MSC films, we chose to use old films that had "isomerized," after sitting on a benchtop for some time. In Ithaca, the humidity during the summer months causes this to occur rapidly, but researchers can promote this to occur using previously published methods (*44*). These films were used instead of the as-made CdS MSCs because the luminescence from these films is much narrower, making the data easier to interpret (*44*).

**Mueller Matrix Polarimetry**

Mueller Matrix Polarimetry (MMP) spectra and maps were collected at the B23 beamline at Diamond Light Source Ltd (beamtime reference ID: SM32994) (*47, 48*). Through MMP, we can resolve the Mueller matrix of our samples to exactly describe the polarization and intensity of the emergent light beam given any incident light beam. The wavelength of the incident light was controlled by a double grating subtractive monochromator and typically swept from 400 – 250 nm with 1 nm steps. The polarization states were modulated using 4 photoelastic modulators (PEMs), two before the sample and two after the sample, operating at different frequencies and rotated at different angles (*15*). The beam was focused to 50 μm x 50 μm pixel size using a 10X microscope UV objective. Using B23 beamline controls and these capabilities enables the measurement of the full, 16 element Mueller Matrix.

The measured Mueller Matrices were converted to the 6 polarization effects (LD, LD`, LB, LB`, CD, and CB) through analytic inversion (*38*). Because the nanocluster films are anisotropic with both chiral and linear polarization-dependent absorption and retardation, the measured Mueller matrix is complex and lacks typical symmetries (*43*). For additional details on the analytical inversion, see our previous work in *Science* (*31*).

**Scanning Electron Microscopy**

Scanning electron microscope (SEM) images were collected on a Zeiss Gemini 500 microscope. Samples were self-assembled on glass slides, as described above. Then, excess glass was trimmed using a glass cutter and the sample was mounted on an SEM sample stub with carbon tape. To reduce charging, the samples were sputter coated with a thin layer of Au/Pd. Then, a small drop of silver paste was dropped on the edge of a sample and allowed to dry, before being grounded to the stub with copper tape. Ultra-low accelerating voltages (1 keV) were also used to limit charging. Samples were imaged at a working distance close to 4 mm.

**Numerical Modeling**

The expressions for the observed LD were written as functions in a Python environment (Python Software Foundation, https://www.python.org/). Using numerical methods, these functions were iterated over the parameter spaces considered. To simulate the polarized optical effects, Lorentzian ($\chi(g,\lambda)$) and Gaussian ($\gamma(g,\lambda)$) functions were used:

$$\chi(g,\lambda) = A\frac{g^2}{(\lambda - \lambda_0)^2 + g^2}$$

$$\gamma(g,\lambda) = Ae^{-\left(\frac{\lambda-\lambda_0}{g}\right)^2}$$

where $\lambda$ and $\lambda_0$ correspond to the wavelength and peak position (typically $\lambda_0 = 324$ nm), $g$ is a broadening constant that was varied, and $A$ is an amplitude constant that was also varied.

To model linear dichroism (LD) and its prime counterpart (LD`), the functions were used as-is. To model circular dichroism (CD) two functions of opposite sign, equal intensity, and equal offset from the absorbance peak position were summed together to create a bisignate lineshape, characteristic of exciton coupling. To model the birefringent counterparts (LB, LB`, and CB) the functions generated above were Hilbert transformed using the Scipy Fast Fourier Transform package (*49*). The absorbance was computed from the polarized optical effects by summing together the squares of the dichroic optical effects and taking their square root. To model a measurement through the back face, the sign of the linear prime effects is flipped.

The linearly-polarized luminescence was modeled using an asymmetric Lorentzian: two Lorentzians were summed together with their peaks at different energies, in order to replicate the low-energy tail that is observed in luminescence measurements of CdS MSCs (*29, 44*).

Notably, there are typically multiple combinations of the excitonic effects that can replicate the lineshapes observed in a LD measurement (**Figure 4**). For this reason, we do urge some caution in the use of this method to extract the characteristic excitonic LD, although reasonable values can be estimated based on simaltaneous CD and LD measurements at different angles. The best method to measure, account, and correct for these artifacts is Mueller matrix polarimetry (MMP).

## Supplementary Text:

### Derivation for Measured LD Signal on CD spectrometers

Light propagating in the $z$-direction, incident on a sample is described by the Stokes vector, $S$:

$$S = \begin{bmatrix} I \\ Q \\ U \\ V \end{bmatrix} = \begin{bmatrix} I_x + I_y \\ I_x - I_y \\ I_{45°} - I_{-45°} \\ I_R - I_L \end{bmatrix} \tag{1}$$

where $I$ is the total intensity of the incident light, $Q$ is the difference in intensity between the $x$- and $y$-axis polarized light with the $x$-axis is taken as the horizontal, $U$ is the difference in intensity between the ±45°-polarized light, and $V$ is the difference in intensities between the right- and left-circularly polarized light.

In CD spectrometers, the polarization of the incident light is generated with a photoelastic modulator (PEM) oriented at 45° (**Fig. S1**). The Stokes vector for this light is

$$S_{in} = \begin{bmatrix} 1 \\ \cos(\delta) \\ 0 \\ \sin(\delta) \end{bmatrix} \qquad (2)$$

where $\delta$ is the time-dependent retardation introduced by the PEM and takes the form, $\delta = A_0 \sin(\omega t + \varphi) + \alpha$. $A_0$ is the amplitude of modulation, $\omega$ is the frequency of the modulator, and $\varphi$ is the phase. The term $\alpha$ is a static retardation in the optical element independent of the dynamic retardation, arising from residual strain in the modulator material. This Stokes vector oscillates between linearly-polarized and circularly-polarized light, enabling the measurement of both CD and LD concurrently (**Fig. S1**). To measure the CD, the instrument locks into the oscillating frequency of the PEM, $\omega$, and to measure the LD, the instrument locks into a harmonic of this oscillation frequency, $2\omega$ (**Fig. S1**).

The incident light from the PEM ($S_{in}$) passes through and is transformed by the sample, and emerges in a new polarization state ($S_{out}$) (**Fig. S1**). The Mueller matrix, $M_z$, is a matrix that describes how a medium transforms the polarization state of the light by relating the incoming and outgoing Stokes vectors:

$$S_{out} = M_z S_{in} \qquad (3)$$

A Mueller matrix can be treated with the lamellar approximation, where the medium is broken up into infinitesimally thin slabs. These slabs are known as differential Mueller matrices, $m$, and they are related to the total Mueller matrix through a spatial derivative. For a non-depolarizing, homogeneous sample, the total Mueller matrix of a sample is related to its differential Mueller matrices through an exponential:

$$M_z = e^{mz} \qquad (4)$$

The differential Mueller matrix, which describes the interaction of light with an infinitesimally thin slab of a sample, is populated with the primary optical effects: absorption ($A$), linear dichroism ($LD$), linear birefringence ($LB$), their prime counterparts measured at 45° ($LD`$, $LB`$), circular dichroism ($CD$), and circular birefringence ($CB$). The differential Mueller matrix takes the form:

$$mz = z \begin{bmatrix} a & -ld & -ld` & cd \\ -ld & a & cb & lb` \\ -ld` & -cb & a & -lb \\ cd & -lb` & lb & a \end{bmatrix} = \begin{bmatrix} A & -LD & -LD` & CD \\ -LD & A & CB & LB` \\ -LD` & -CB & A & -LB \\ CD & -LB` & LB & A \end{bmatrix}$$

$$= A\mathbb{1} + \begin{bmatrix} 0 & -LD & -LD` & CD \\ -LD & 0 & CB & LB` \\ -LD` & -CB & 0 & -LB \\ CD & -LB` & LB & 0 \end{bmatrix} = A\mathbb{1} + F$$

where $\mathbb{1}$ is the identity matrix. The lowercase optical effect notations are expressions for the same optical effects but expressed in per unit length notation ($ld \equiv \frac{LD}{z}$) (7, 37). When the sample is homogeneous throughout $z$, the matrices and resulting expressions become $z$-independent (7, 37).

The exponential form of the Mueller matrix, $M_z = e^{mz}$, is convenient because it can be approximated with a Taylor expansion, enabling one to solve equation (3) through matrix multiplication alone. A second order Taylor expansion is typically used, and the exponential form of the Mueller matrix becomes:

$$M_z = e^{mz} = e^{A\mathbb{1}}e^F = e^{A\mathbb{1}}\left[\mathbb{1} + F + \frac{1}{2!}F^2 + \cdots\right] \quad (5)$$

Because a sample can be rotated azimuthally ($\theta$) about the light propagation direction, rotation matrices are applied to the Mueller matrix, yielding an angular dependent Mueller matrix, $M'(\theta)$:

$$S_{out} = R(-\theta)M_zR(\theta)S_{in} = M'(\theta)S_{in} \quad (6)$$

Modern photodetectors only measure the total intensity of the light, which corresponds to the first component of the outgoing Stokes vector. Thus, we take only the first component of the Stokes vector, $I_{detect}$. This signal is composed of both time-varying (AC) and constant (DC) components. The oscillation frequencies of the time-varying components are determined by the driving voltage that is applied to the PEM, and there are components at this driving frequency, $\omega$, as well as all of its harmonics, $2\omega$, $3\omega$, $4\omega$, and so on. A lock-in amplifier isolates the time-varying components oscillating at $\omega$ and $2\omega$ from the DC component. The remainder of the harmonics are not isolated or used in standard CD spectrometers. The ratio between the $\omega$ component and the DC component is reported as the measured CD, $I_{CD}$, and the ratio between the $2\omega$ component and the DC component is reported as the measured LD, $I_{LD}$.

Depending on the order of the Taylor expansion used to model the exponential form of the Mueller matrix derived through the lamellar approximation, the first component of the Stokes vector, $I_{detect}$, will take on a different form:

$$I_{detect}(1) = e^A\{1 + CD\sin\delta + (LD`\sin 2\theta - LD\cos 2\theta)\cos\delta\}$$

$$I_{detect}(2) = e^A \left\{ 1 + \frac{1}{2}(CD^2 + LD^2 + LD^{`2}) + \left(CD + \frac{1}{2}(LBLD` - LDLB`)\right) \sin \delta \right.$$
$$\left. + \left(\left(LD` - \frac{1}{2}(CDLB - CBLD)\right) \sin 2\theta - \left(LD - \frac{1}{2}(CBLD` - CDLB`)\right) \cos 2\theta \right) \cos \delta \right\}$$

$$I_{detect}(3)$$
$$= e^A \left\{ 1 + \frac{1}{2}(CD^2 + LD^2 + LD^{`2}) \right.$$
$$+ \frac{1}{6}(CD(LBLD` - LDLB`) - LD(CBLD` - CDLB`) - LD`(CDLB - CBLD))$$
$$+ \left(CD + \frac{1}{2}(LBLD` - LDLB`)\right.$$
$$+ \frac{1}{6}(CD(CD^2 + LD^2 + LD^{`2}) - LB(CDLB - CBLD) + LB`(CBLD` - CDLB`))\bigg) \sin \delta$$
$$+ \left(\left(LD` - \frac{1}{2}(CDLB - CBLD) - \frac{1}{6}(CB(CBLD` - CDLB`) + LB(LBLD` - LDLB`) - LD`(CD^2 + LD^2 + LD^{`2}))\right) \sin 2\theta \right.$$
$$- \left(LD - \frac{1}{2}(CBLD` - CDLB`) \right.$$
$$\left. \left. + \frac{1}{6}(CB(CDLB - CBLD) + LB(LBLD` - LDLB`) + LD(CD^2 + LD^2 + LD^{`2}))\right) \cos 2\theta \right) \cos \delta \right\}$$

These expressions, however, each take on a common form:

$$I_{detect} = e^A \{C_0 + C_{CD} \sin \delta + C_{LD} \cos \delta\}$$

where the values for each of the pre-factors, $C_0$, $C_{CD}$, and $C_{LD}$, depend on the order of the Taylor expansion. The term $\delta$ is the time-dependent retardation introduced by the PEM, written as

$$\delta(t) = A_0 \sin(\omega t + \varphi) + \alpha$$

where $A_0$ is the amplitude of modulation, $\omega$ is the frequency of the modulator, and $\varphi$ is the phase. The term $\alpha$ is a static retardation in the optical element independent of the dynamic retardation, arising from residual strain in the modulator material. This term is very small for modern PEMs.

The time-varying detector signal components, $\sin \delta$ and $\cos \delta$, can be expanded in an infinite series of integer Bessel functions through the sum and difference trigonometric identities followed by the Jacobi-Anger expansion

$$\sin \delta = \sin(A_0 \sin(\omega t + \varphi) + \alpha) = \sin(A_0 \sin(\omega t + \varphi)) \cos \alpha + \cos(A_0 \sin(\omega t + \varphi)) \sin \alpha$$

$$\sin\delta = 2\cos\alpha\left[\sum_{n=1}^{\infty} J_{2n-1}(A_0)\sin((2n-1)(\omega t + \varphi))\right]$$
$$+ \sin\alpha\left[J_0(A_0) + 2\sum_{n=1}^{\infty} J_{2n}(A_0)\cos(2n(\omega t + \varphi))\right]$$

$$\cos\delta = \cos(A_0\sin(\omega t + \varphi) + \alpha) = \cos(A_0\sin(\omega t + \varphi))\cos\alpha - \sin(A_0\sin(\omega t + \varphi))\sin\alpha$$

$$\cos\delta = \cos\alpha\left[J_0(A_0) + 2\sum_{n=1}^{\infty} J_{2n}(A_0)\cos(2n(\omega t + \varphi))\right]$$
$$- 2\sin\alpha\left[\sum_{n=1}^{\infty} J_{2n-1}(A_0)\sin((2n-1)(\omega t + \varphi))\right]$$

In this work, we will only use these expansions out to order $n = 1$, since the lock-ins used in CD spectrometers only use the $\omega$ and $2\omega$ components. In techniques where lock-ins are not used such as Mueller matrix polarimetry, however, the higher order harmonics are used as well to improve the quality of and confidence in the collected data. The expansions above for the time-varying detector signal components can now be substituted into $I_{detect}$, expanded to order $n = 1$ and written as

$$I_{detect} = e^A\{C_0$$
$$+ C_{CD}\left[2\cos\alpha J_1(A_0)\sin(\omega t + \varphi) + \sin\alpha\left[J_0(A_0) + 2J_2(A_0)\cos(2(\omega t + \varphi))\right]\right]$$
$$+ C_{LD}\left[\cos\alpha\left[J_0(A_0) + 2J_2(A_0)\cos(2(\omega t + \varphi))\right] - 2\sin\alpha J_1(A_0)\sin(\omega t + \varphi)\right]\}$$

To convert these signals into the CD and LD that are reported on a CD spectrometer, a lock-in-amplifier is used to break these up into the AC and DC components. The AC component is comprised of components operating at $\omega$ and $2\omega$. The CD signal is reported as the ratio between the $\omega$ component and the DC component, whereas the LD signal is reported as the ratio between the $2\omega$ component and the DC component. Each can be written as

$$CD_{measured} = \frac{AC_\omega}{DC} = \frac{2C_{CD}\cos\alpha J_1(A_0)\sin(\omega t + \varphi) - 2C_{LD}\sin\alpha J_1(A_0)\sin(\omega t + \varphi)}{C_0 + C_{CD}J_0(A_0)\sin\alpha + C_{LD}J_0(A_0)\cos\alpha}$$

$$LD_{measured} = \frac{AC_{2\omega}}{DC} = \frac{2C_{CD}\sin\alpha J_2(A_0)\cos(2(\omega t + \varphi)) + 2C_{LD}\cos\alpha J_2(A_0)\cos(2(\omega t + \varphi))}{C_0 + C_{CD}J_0(A_0)\sin\alpha + C_{LD}J_0(A_0)\cos\alpha}$$

These expressions can be further simplified by using the small angle identities (recall, $\alpha$, the static retardation of the PEM, is very small) and by ignoring the denominator because it is a constant. The expressions for the CD and LD signals are now written as

$$CD_{measured} = (C_{CD} - \alpha C_{LD})(2J_1(A_0)\sin(\omega t + \varphi))$$

$$LD_{measured} = (\alpha C_{CD} + C_{LD})(2J_2(A_0)\cos(2(\omega t + \varphi)))$$

where the constants, $C_{CD}$ and $C_{LD}$ depend on the order of the Taylor expansion. The Bessel functions and time-varying sine and cosine functions will not impact the measured signal: the Bessel function is a constant, determined by the amplitude of the PEM modulation, $A_0$, and the time-varying signal is the reference for the lock-in. Thus, we discard these terms and write the measured CD and LD as

$$CD_{measured} = C_{CD} - \alpha C_{LD}$$

$$LD_{measured} = \alpha C_{CD} + C_{LD}$$

where the dominant terms for the measured CD and LD are the constants $C_{CD}$ and $C_{LD}$, respectively. In the case of an ideal PEM, where the residual static birefringence is zero, $\alpha = 0$, the expressions reduce even further to

$$CD_{measured} = C_{CD}$$

$$LD_{measured} = C_{LD}.$$

**Tables of constants used in the above expressions**

**Table S1**

| Taylor Expansion | $C_0$ |
|---|---|
| 1st | 1 |
| 2nd | $1 + \frac{1}{2}(CD^2 + LD^2 + LD`^2)$ |
| 3rd | $1 + \frac{1}{2}(CD^2 + LD^2 + LD`^2) + \frac{1}{6}(CD(LBLD` - LDLB`) - LD(CBLD` - CDLB`) - LD`(CDLB - CBLD))$ |

**Table S2**

| Taylor Expansion | $C_{CD}$ |
|---|---|
| 1st | $CD$ |
| 2nd | $CD + \frac{1}{2}(LBLD` - LDLB`)$ |
| 3rd | $CD + \frac{1}{2}(LBLD` - LDLB`) + \frac{1}{6}(CD(CD^2 + LD^2 + LD`^2) - LB(CDLB - CBLD) + LB`(CBLD` - CDLB`))$ |

**Table S3**

| Taylor Expansion | $C_{LD}$ |
|---|---|
| 1st | $LD\` \sin 2\theta - LD \cos 2\theta$ |
| 2nd | $\left(LD\` - \frac{1}{2}(CDLB - CBLD)\right)\sin 2\theta - \left(LD - \frac{1}{2}(CBLD\` - CDLB\`)\right)\cos 2\theta$ |
| 3rd | $\left(LD\` - \frac{1}{2}(CDLB - CBLD) - \frac{1}{6}(CB(CBLD\` - CDLB\`) + LB(LBLD\` - LDLB\`) - LD\`(CD^2 + LD^2 + LD\`^2))\right)\sin 2\theta - \left(LD - \frac{1}{2}(CBLD\` - CDLB\`) + \frac{1}{6}(CB(CDLB\` - CBLD\`) + LB(LBLD\` - LDLB\`) + LD(CD^2 + LD^2 + LD\`^2))\right)\cos 2\theta$ |

### Derivation of Expressions to Isolate Nonreciprocal LD

In order to isolate the nonreciprocal LD component from experimental measurements, we derived equations (3) and (4). From equation (2b), the measured LD can be expressed as

$$LD_{measured} = LD_{natural} + LD_{CDLB}$$

More specifically, we consider two measurements, through the front and back of the sample:

$$LD_{measured,Front} = LD_{natural} + LD_{CDLB,Front}$$

$$LD_{measured,Back} = LD_{natural} + LD_{CDLB,Back}$$

Where $I_{LD,Front}$ is the measured LD intensity in front illumination conditions, and $I_{LD,Back}$ is the measured LD intensity under back illumination.

Based on the fact that the nonreciprocal portion inverts upon sample flipping, we assume

$$LD_{CDLB,Front} = -LD_{CDLB,Back}$$

Plugging in this value and solving yields equations (3) and (4) in the maintext. Working this out for the 'front' term:

$$LD_{measured,Back} = LD_{natural} + LD_{CDLB,Back} = LD_{natural} - LD_{CDLB,Front}$$

Subtract from the front term

$$LD_{measured,Front} - LD_{measured,Back} = LD_{natural} + LD_{CDLB,Front} - LD_{natural} + LD_{CDLB,Front}$$

$$= 2LD_{CDLB,Front}$$

Which gives the nonreciprocal contribution in front illumination

$$LD_{CDLB,Front} = LD_{nonreciprocal,Front} = \frac{1}{2}(LD_{measured,Front} - LD_{measured,Back}) \quad (3)$$

And equally

$$LD_{CDLB,Back} = \frac{1}{2}(LD_{measured,Back} - LD_{measured,Front}) \qquad (4)$$

### Derivation of Linearly Polarized Luminescence Measurements

In order to derive the expressions for linearly-polarized luminescence measurements, we use a modified Stokes-Mueller formalism that connects absorption processes to emission processes (*36*). The key expression is

$$F = \Phi TM$$

where $F$ is the fluorescence matrix, $\Phi$ is the fluorescence quantum yield, T is the 4x4 absorption-fluorescence transformation matrix, and $M$ is the 4x4 expanded Mueller matrix (*36*). Plugging in the second-order expanded Mueller matrix enables the computation of the fluorescence matrix. In this work, we consider the $F_{00}$ matrix element, which is the fluorescence from the sample upon unpolarized excitation. This full expression comes out as

$$F_{00} = \frac{\Phi}{2}[T_{00}(2 + LD^2 + LD`^2 + CD^2) + T_{01}(2LD + CBLD` - CDLB`) \\ + T_{02}(2LD` + CDLB - CBLD) + T_{03}(2CD + LDLB` - LBLD`)]$$

From this expression, we can see that all of the matrix elements from the first row of the expanded Mueller matrix contribute to the fluorescence upon unpolarized excitation. This expression would accurately describe the nonreciprocal, circularly-polarized luminescence observed in LDLB samples (*12*). For simplicity, we set $T_{00} = T_{02} = T_{03} = 0$ in our model, although in a real sample this is likely not the case.

### Previous Commentary on Nonreciprocal LD Terms

Previous authors have derived expressions for LD measurements, revealing the same CDLB` and CBLD` terms derived in this work. However, these terms have largely been considered negligible. To our knowledge, the first discussion of these terms came from Jensen, Schellman, and Troxell in 1978 (*25*):

"…in an attempt to measure LD, instrumental circular birefringence will couple with linear dichroism of the sample to give a spurious signal which, however, is unimportant, as such circular anisotropics are rare and easy to avoid."

Schellman and Jensen revisited these terms in a review paper in 1987, writing(*26*):

"*Certain types of measurements are now standard and direct. CD and CB can be easily measured in isotropic samples. LD and LB are easily measured in achiral samples and*

*usually in chiral samples as well if the linear polarization effects are very much stronger than the circular polarization effects, which is usually the case. Difficulties arise in attempting to measure circular anisotropies in the presence of strong linear anisotropies or whenever circular and linear anisotropies are of the same order of magnitude."*

It should be noted that there is more extensive discussion on the terms elsewhere in their review, too. Shindo uncovered similar terms that interfere with LD measurements in 1990, writing (*19*):

*"The first term is the contribution of LD to a measured spectrum and dependent upon the rotation of the sample. The second term is due to the coupling of the mixed anisotropies of the sample with nonideal characteristics of the modulator, and independent of the sample rotation. The first term is at least 100 times larger than the second term, and the CD contribution to the measured spectrum is negligible. Thus the Signal CH2 is also not a linear combination of CD and LD signals."*

To the best of our knowledge, there is no extensively novel discussion about these terms contributing to an LD measurement since these works. They have appeared since then, for instance in the Ph.D. thesis of one of the authors of this work, Oriol Arteaga, who made similar assumptions about small values for the chiroptic effects (*50*).

**On the Magnitudes of the Optical Effects Needed for Nonreciprocal LD**

As discussed in the maintext and in the previous section, the nonreciprocal LD terms have been derived by previous researchers, but ignored under the assumption that the circular anisotropy (CD, CB) is orders of magnitude weaker than the linear anisotropy (LD, LB, LD`, LB`). The CdS MSCs are uniquely suited to observe this effect, courtesy of their exceptionally strong circular anisotropy and their highly aligned transition dipole moments. In this work, we observed nonreciprocal LD in CdS, CdSe, and CdTe films. These films possess varying degrees of circular and linear anisotropy.

Beginning with the CdS, these films routinely present CD dissymmetry factors approaching and exceeding unity, and reduced LD exceeding 0.6. Because these effects are so large and on the same order of magnitude, nonreciprocal LD is easily observed if the film is oriented so that the dipoles are along the diagonal. At this orientation, the natural LD is close to 0, but the LD` effects are strong. Because of this, the CDLB` and CBLD` terms are significant and measurable.

Now, we turn our attention to CdSe and CdTe films. The CdSe film presents CD on the order of 0.01 ΔA, and natural LD on the order of 0.1 ΔA. These values are close enough to yield a noticeable nonreciprocal LD signal. The CdTe film presents CD on the order of 0.002 ΔA, and natural LD on the order of 0.02 ΔA. Again, these values are close enough in magnitude to measure a noticeable nonreciprocal LD signal

Several modeled conditions are presented to demonstrate the impact of varying magnitudes of the optical effects on measured spectra (**Fig. S30, S31**).

**Slabs vs. Differential Matrices**

In the past several years, several groups have presented models to intuitively understand the LDLB interference. Some of these models, however, are misleading in that they treat the LDLB interference using two separate slabs, one possessing LB` followed by another possessing LD (**Fig. S27a**)(*22*). While this pair of slabs will, in this order, preferentially absorb one of left- or right-circularly polarized light, it will not when illuminated in the opposite direction (**Fig. S27a**). Further, it breaks the homogeneous medium assumption, which was made to relate the differential Mueller matrix, $m$, to the macroscopic Mueller matrix of the medium, $M_z$, through an exponential. In a real sample possessing both linear and linear prime effects, the light experiences the birefringence and dichroism concurrently, over and over as it propagates through the medium, altering its polarization state as it propagates. A similar picture can be presented for LD measurements, but suffers from the same pitfalls (**Fig. S27b**).

Providing more detail on the CB + LD` stacked slabs, the CB slab will rotate each of the incident vertical and horizontal polarization states, and whichever of these now-rotated polarization states more closely align with the absorbers in the LD` slab will be absorbed to a greater extent, yielding an LD signal. This picture of stacked elements, though, suffers from the same pitfalls as the LDLB effect outlined above (**S27a**).

While on this topic, we also remind the reader that, in the context of **Figure 5**, we are displaying differential Mueller matrices. The differential Mueller matrices correspond to infinitesimally thin slices of the homogeneous media. As the light propagates through many of these matrices/slices, its polarization state is continuously modulated.

**On the various LD spectral lineshapes observed:**

We present in the supplemental materials a handful of LD spectral lineshapes. We have observed countless different LD spectral lineshapes; these depend on the relative magnitudes of the optical effects within the sample and the sample's orientation with respect to the incident beam. For instance, we observe a spectrum that has a narrow blip in the LD, aligned with the CD peak position (**Fig. S6b**). We hypothesize that this feature is caused by CD interfering with LB`. There are other forms that the artifacts can take (**Fig. S6c**), only strengthening the notion that experimentalists should take extreme care when characterizing optically active samples imbued with linear anisotropy. Regardless of their form, the interplay between the circular effects and the linear prime effects (CDLB` and CBLD`) are the dominant terms that yield these artifacts for LD.

**Polarimetric vs. Lorentz (*i.e.*, electromagnetic) Reciprocity**

Throughout this work, we use the term "nonreciprocal" to describe some effects in circular and linear dichroism spectroscopy. When using this term, we are referring to polarimetric reciprocity, rather than Lorentz reciprocity. Unlike Lorentz reciprocity, polarimetric reciprocity is easily broken by ordinary anisotropic materials (*43*). In a material that preserves polarimetric reciprocity, reversing the direction of light propagation will not change the experiment. In other words, the Stokes vector measured at the output will be the same after reversing the wave vector (*43*).

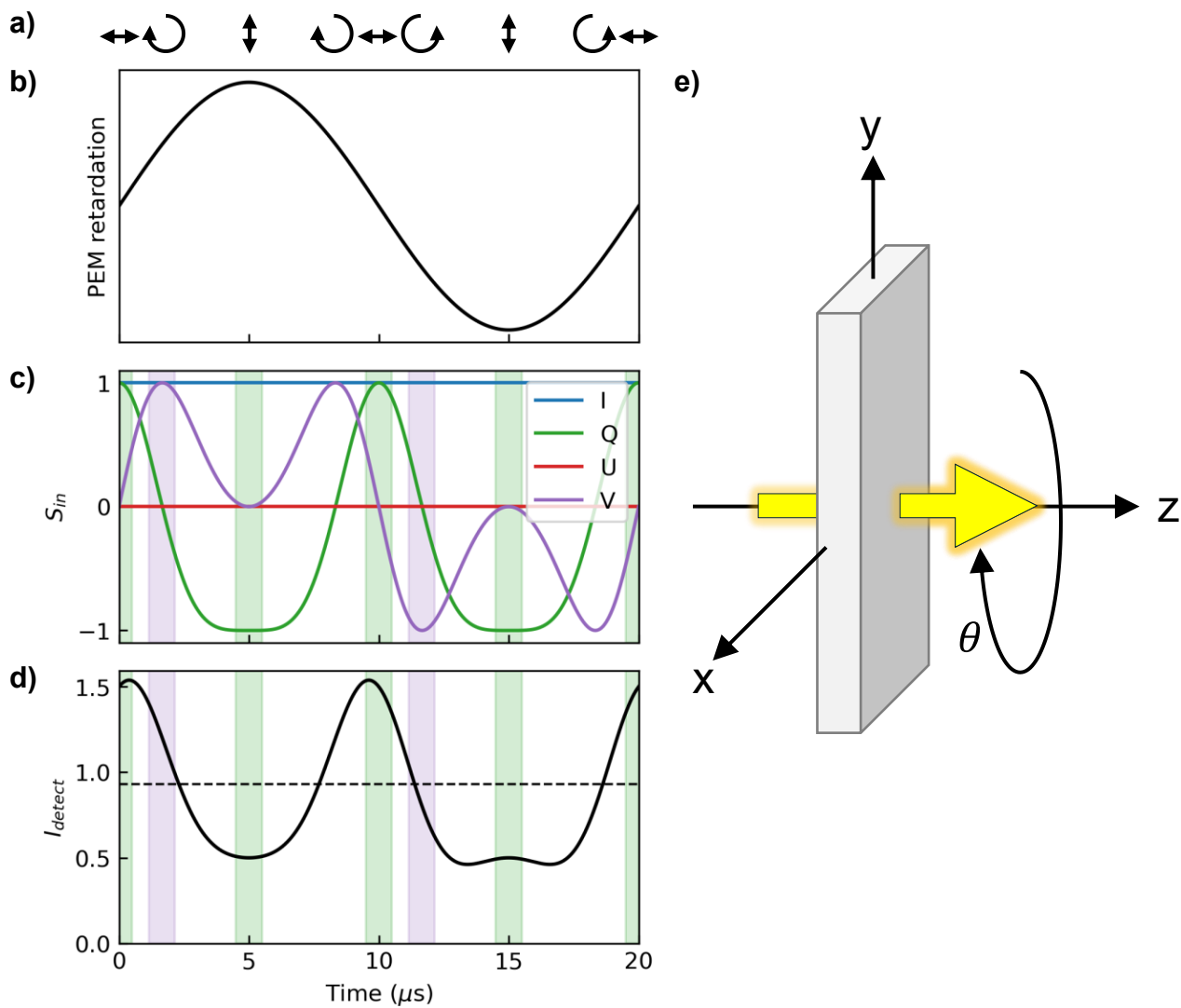

**Figure S1.** Operating principle of CD spectrometers. **a)** Polarization-state of the light exiting the PEM as a function of time. **b)** Retardation of a PEM as a function of time, oscillating between 0, ±¼-wave, and ±½-wave. The PEM is driven by a sinusoidal voltage (see maintext derivation and eqs. (1) and (2)). Values used for plot: $A_0 = 0, \omega = 50000$ kHz, $\alpha = \varphi = 0$ **c)** Values of the incident Stokes vector as a function of time. The I component is normalized to 1. U is 0 at all times, and Q and V oscillate between 0 and ±1. **d)** Detector signal as a function of time. The intensity and lineshape will depend on the chiral and linear anisotropy, if any at all. For an achiral isotropic sample, the detector signal will be DC. For a sample with only CD, the time-varying signal will lock-in exactly to V. For a sample with only LD, the time-varying signal will lock-in exactly to Q. Values used: CD=0.2, LD=0.5 **e)** Schematic showing the coordinate system and definition of the angle, θ, used in the Stokes-Mueller calculus.

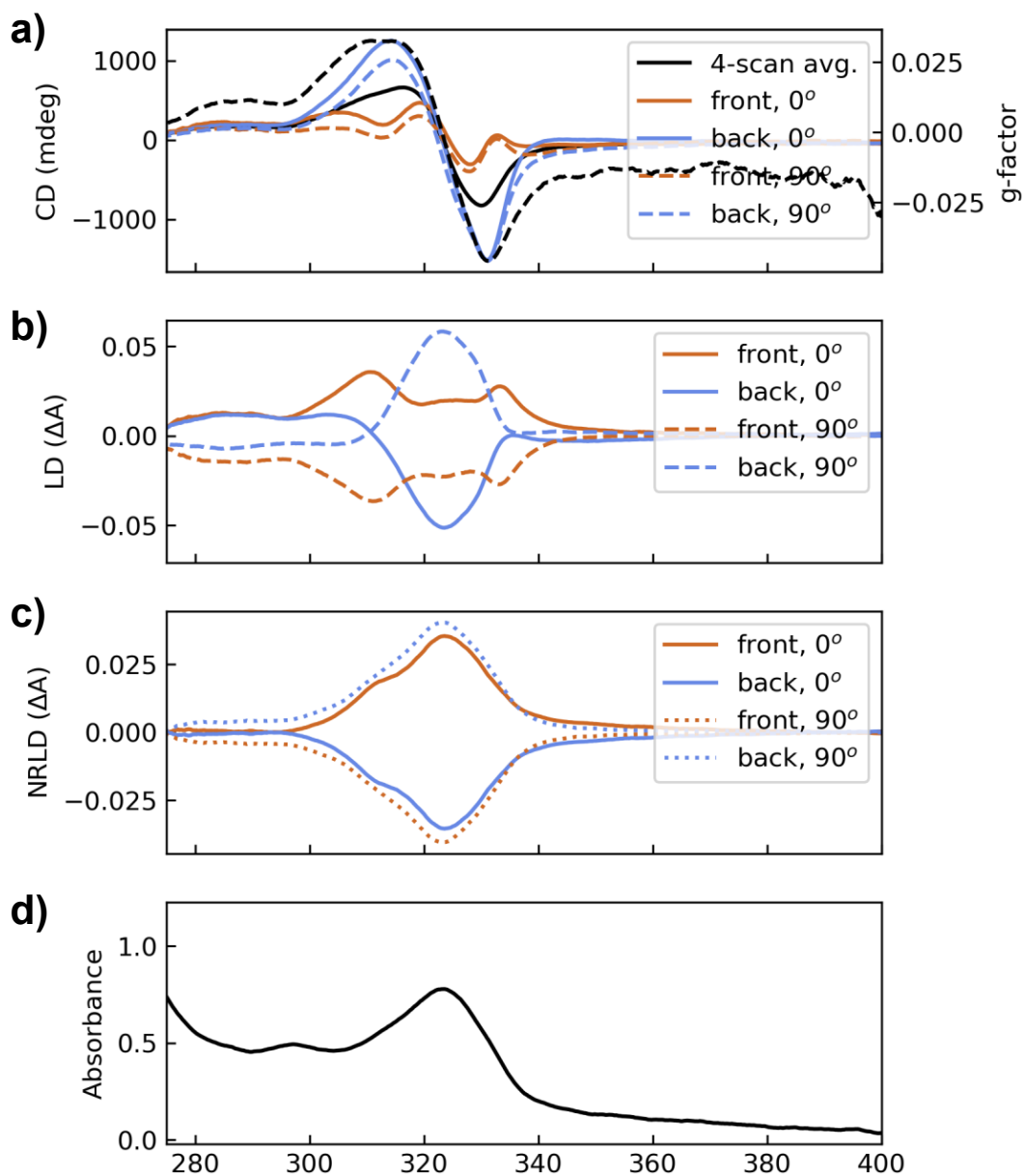

**Figure S2.** Experimental data used in figure 2. **a)** CD (solid black curve) and CD g-factor (dashed black curve) measured from a CdS MSC film. The black spectra are the average of the four spectra collected for this sample, a methodology derived in our group's previous work (Yao, *et al.*, ACS Nano, 2022). The colored spectra are the individual scans at each orientation. **b)** LD spectra collected at each of the four orientations. This is the same data as presented in Figure 2d. **c)** The NRLD components, computed from the spectra shown in panel b, using equations (3) and (4) in the maintext. **d)** Absorbance measured for the CdS film. The absorbance is computed as the average of the four absorbances measured at each orientation.

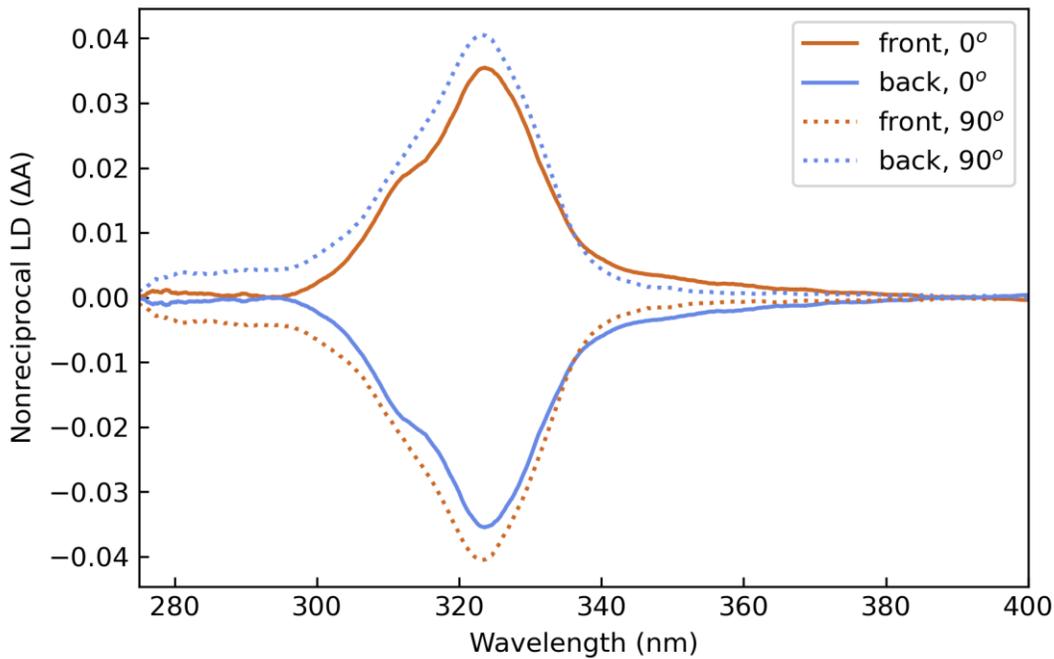

**Figure S3.** NRLD computed in figure 2. The NRLD is computed using equations (3) and (4) in the maintext. Notably, after a 90 degree rotation, the sign of the NRLD component inverts, but the magnitude remains constant. As expected from the Stokes-Mueller theory, the contribution flips sign upon sample flipping.

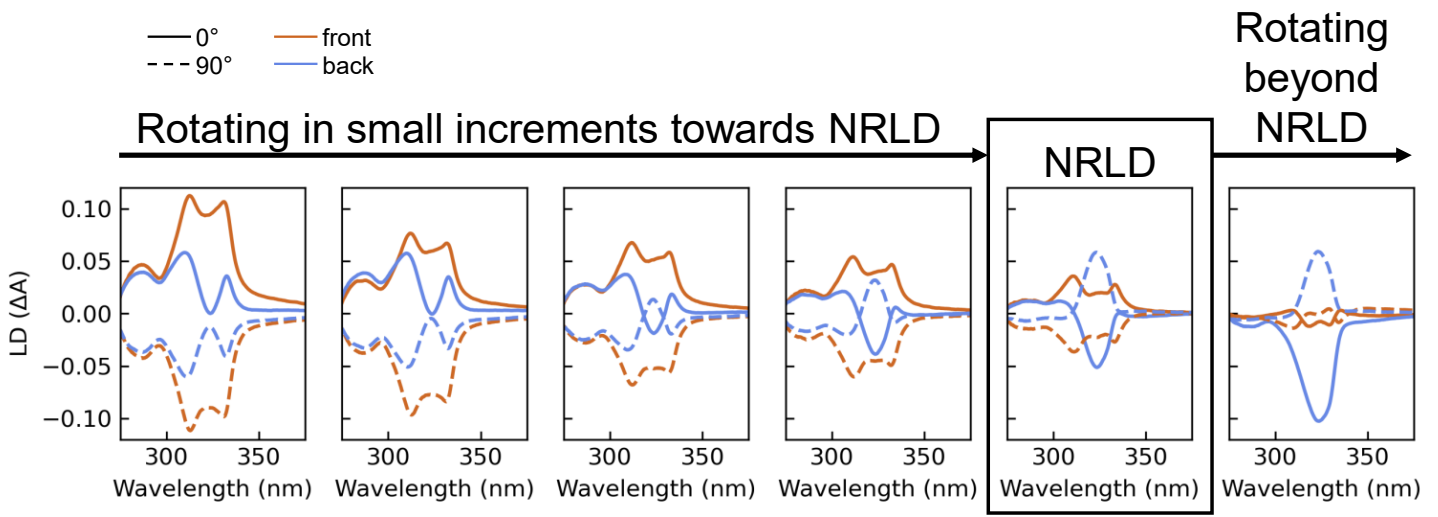

**Figure S4.** LD measured at angles approaching the ideal angle in the CdS MSC film used in Figure 2. The solid curves are measured at 0 degrees, and the dashed curves are measured after a 90-degree rotation. The orange curves correspond to a measurement through the front of the film, and the blue curves correspond to a measurement through the back of the film. The exact angular increments from left-to-right are not exactly known, but they are all within ~5 degrees of the ideal position, which is boxed. At the non-ideal angles, natural LD contribution overwhelms the NRLD contribution. Interestingly, in the right-most measurement, the LD is close to zero when measured through the front and becomes a positive or negative feature when the sample is flipped, depending on the orientation. This is an interesting spectral feature courtesy of NRLD that could be realized and utilized for photonic applications.

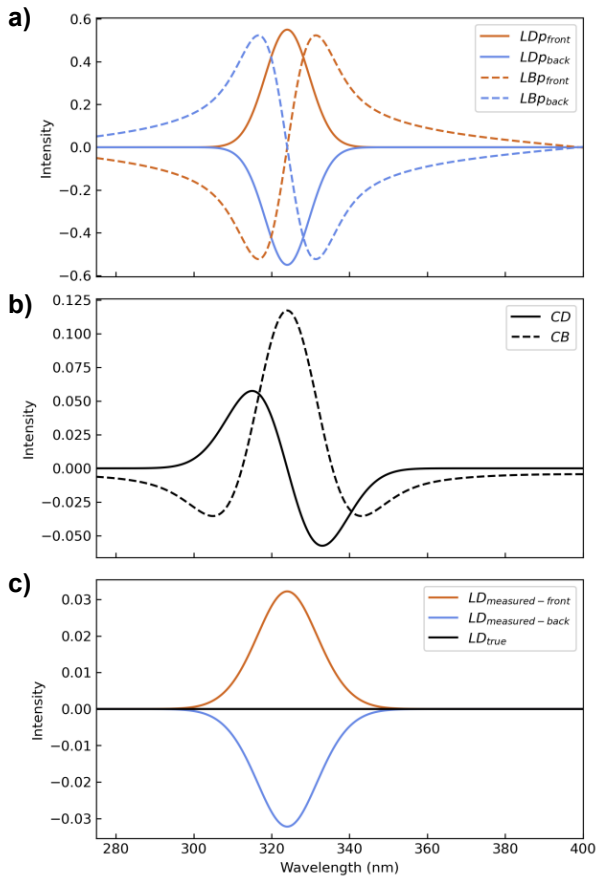

**Figure S5.** Modelled Nonreciprocal LD & LB spectra used in **Figure 2**. **a)** The LD` and LB` used for the front and back faces of the film. **b)** The CD and CB used for both faces of the film (considered rotationally invariant). **c)** The computed nonreciprocal LD, compared to the "true" LD, which was set to 0.

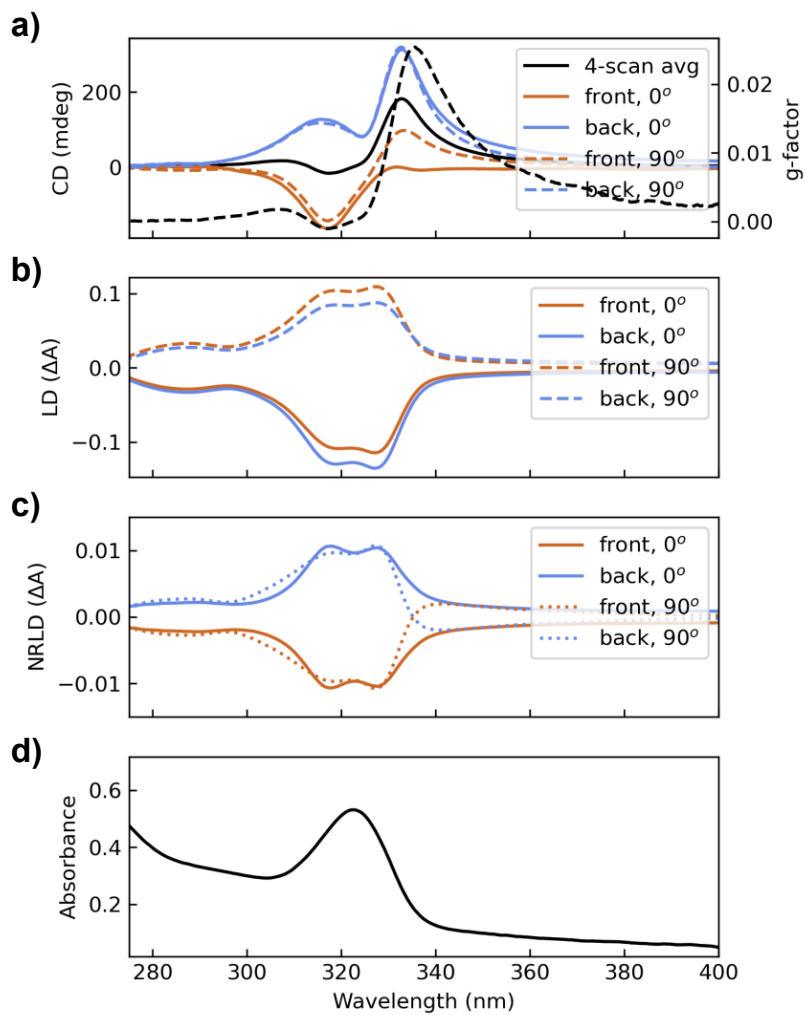

**Figure S6.** Experimental data used in **Figure S7 a)** CD (solid black curve) and CD g-factor (dashed black curve) measured from a CdS MSC film. The black spectra are the average of the four spectra collected for this sample, a methodology derived in our group's previous work (Yao, *et al.*, ACS Nano, 2022). The colored spectra are the individual scans at each orientation. **b)** LD spectra collected at each of the four orientations. This is the same data as presented in Figure 2d. **c)** The NRLD components, computed from the spectra shown in panel b, using equations (3) and (4) in the maintext. **d)** Absorbance measured for the CdS film. The absorbance is computed as the average of the four absorbances measured at each orientation.

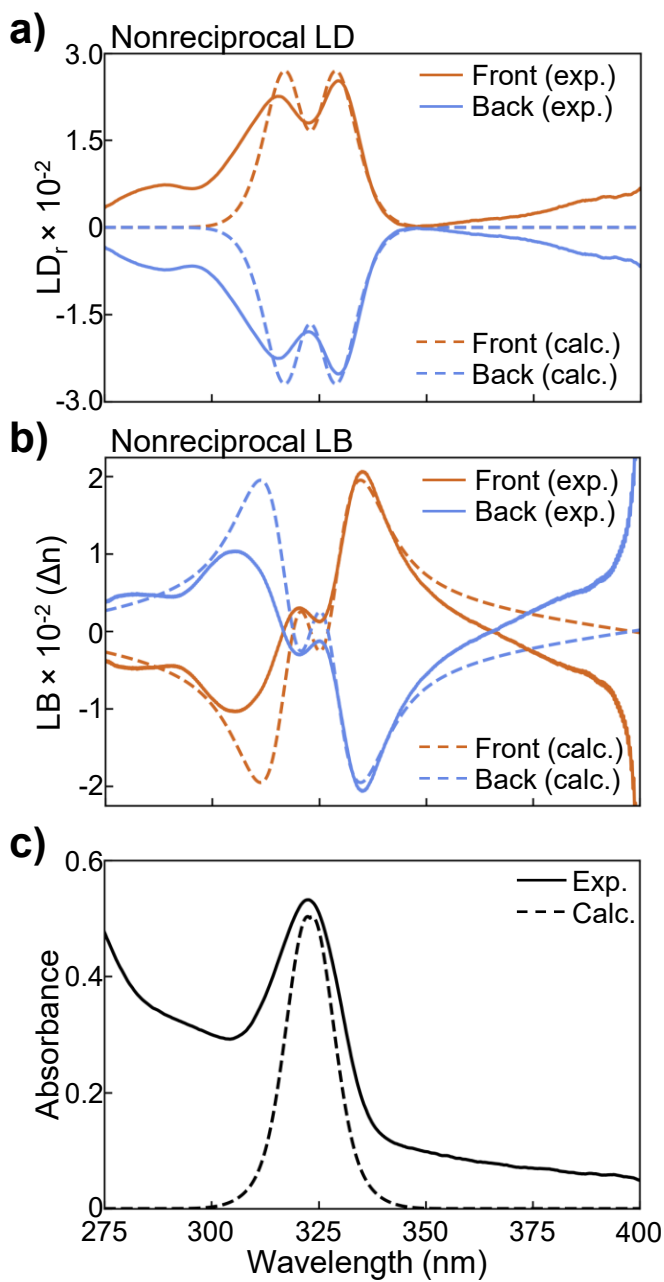

**Figure S7.** Comparison between experiment and model for another dataset. The methodology used in this analysis is the same as in **Figure 2**. **a)** Nonreciprocal LD, isolated using equations (3) and (4) in the maintext. The model (**Figure S8**) achieves strong agreement with the experimental data (**Figure S6**). **b)** Nonreciprocal LB, computed from the data and model in panel **a** through Kramers-Kronig. **c)** The absorbance of the film and model.

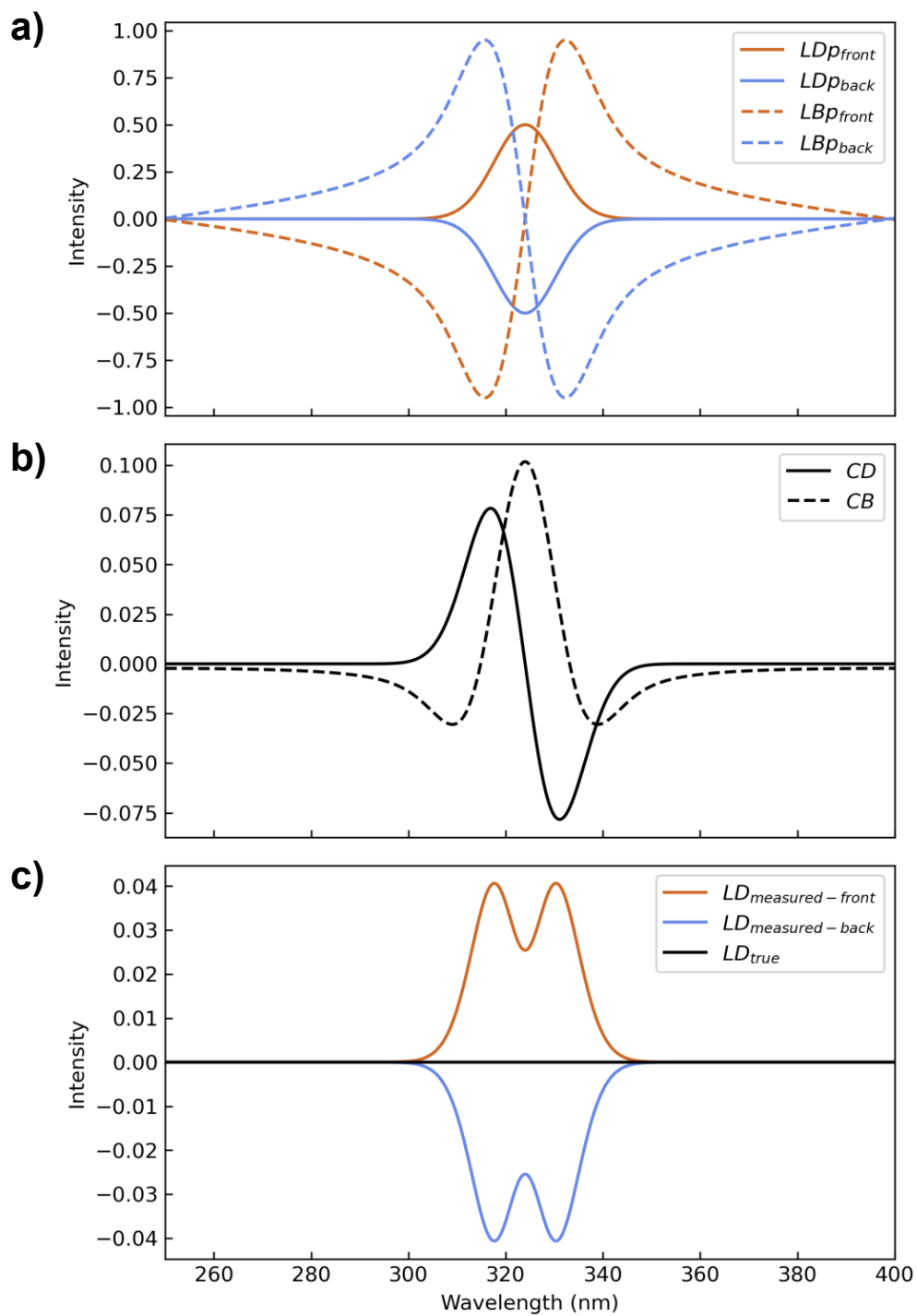

**Figure S8.** Modelled Nonreciprocal LD & LB data used to match experimental data in **Figures S6 and S7**. **a)** The LD` and LB` used for the front and back faces of the film. **b)** The CD and CB used for both faces of the film (considered rotationally invariant). **c)** The computed nonreciprocal LD, compared to the "true" LD, which was set to 0.

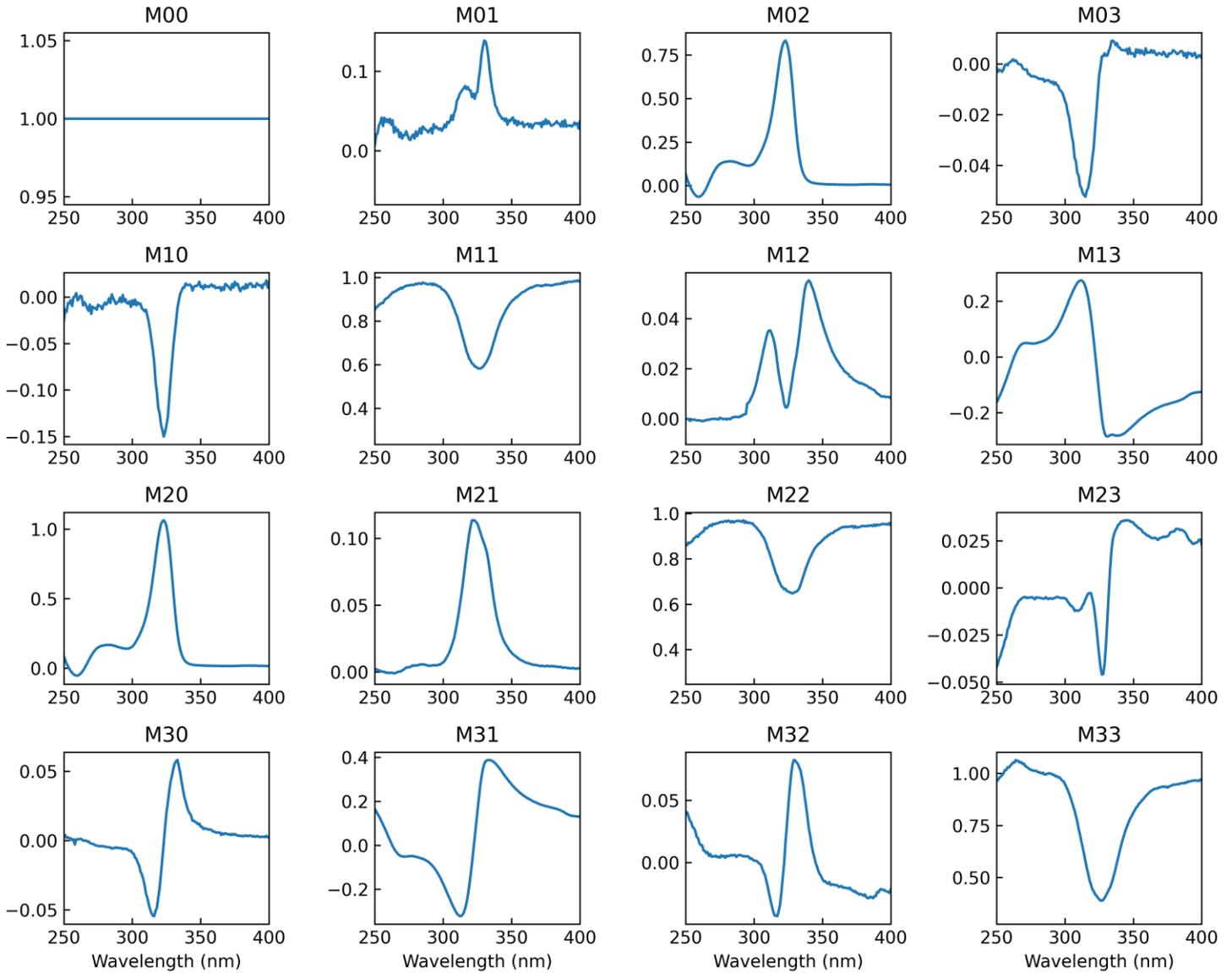

**Figure S9.** Mueller matrix polarimetry (MMP) spectra collected from a CdS MSC film, normalized to the M00 matrix element. This film presents NRLD, evidenced by the sign flip between the M01 and M10 matrix elements. This measurement was made in a position where the LD` is strong and the circular dichroism is strong (g-factor ~ 0.26 from analytical inversion). The LD is weak at this position, because the transitions are oriented along the diagonal.

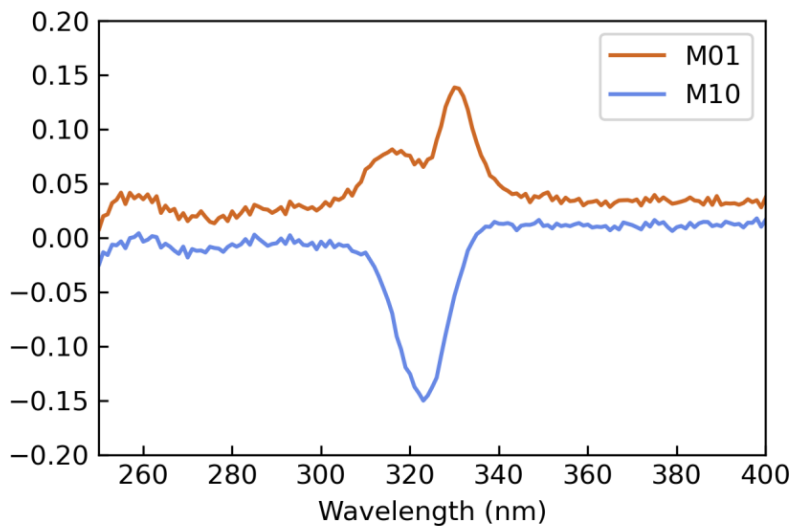

**Figure S10.** Mueller matrix polarimetry (MMP) spectra collected from a CdS MSC film, normalized to the M00 matrix element. Here, we highlight the M01 and M10 matrix elements, which are of opposite sign and equal magnitude, demonstrating the NRLD effect. In samples dominated by natural LD, these two matrix elements are of the same sign and magnitude.

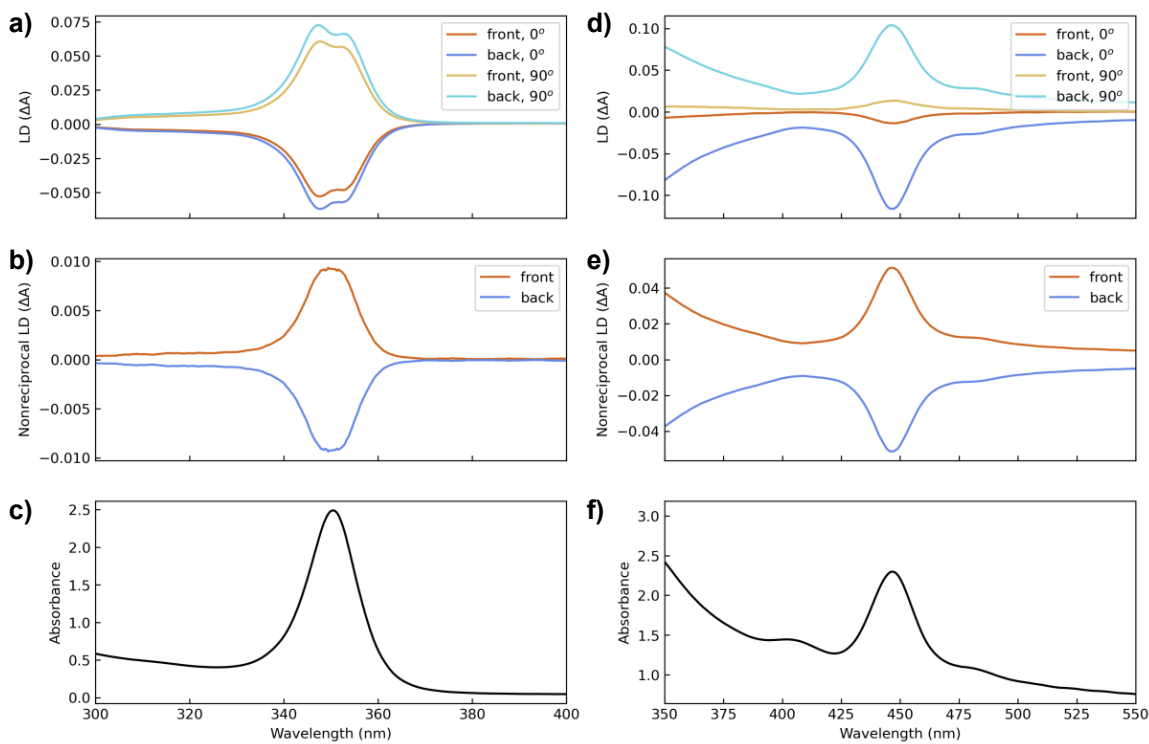

**Figure S11.** CdSe and CdTe linear dichroism spectra. **a-c)** LD, NRLD, and Absorbance of a CdSe MSC film, displaying moderate NRLD. **d-f)** LD, NRLD, and Absorbance of a CdTe MSC film, displaying strong NRLD.

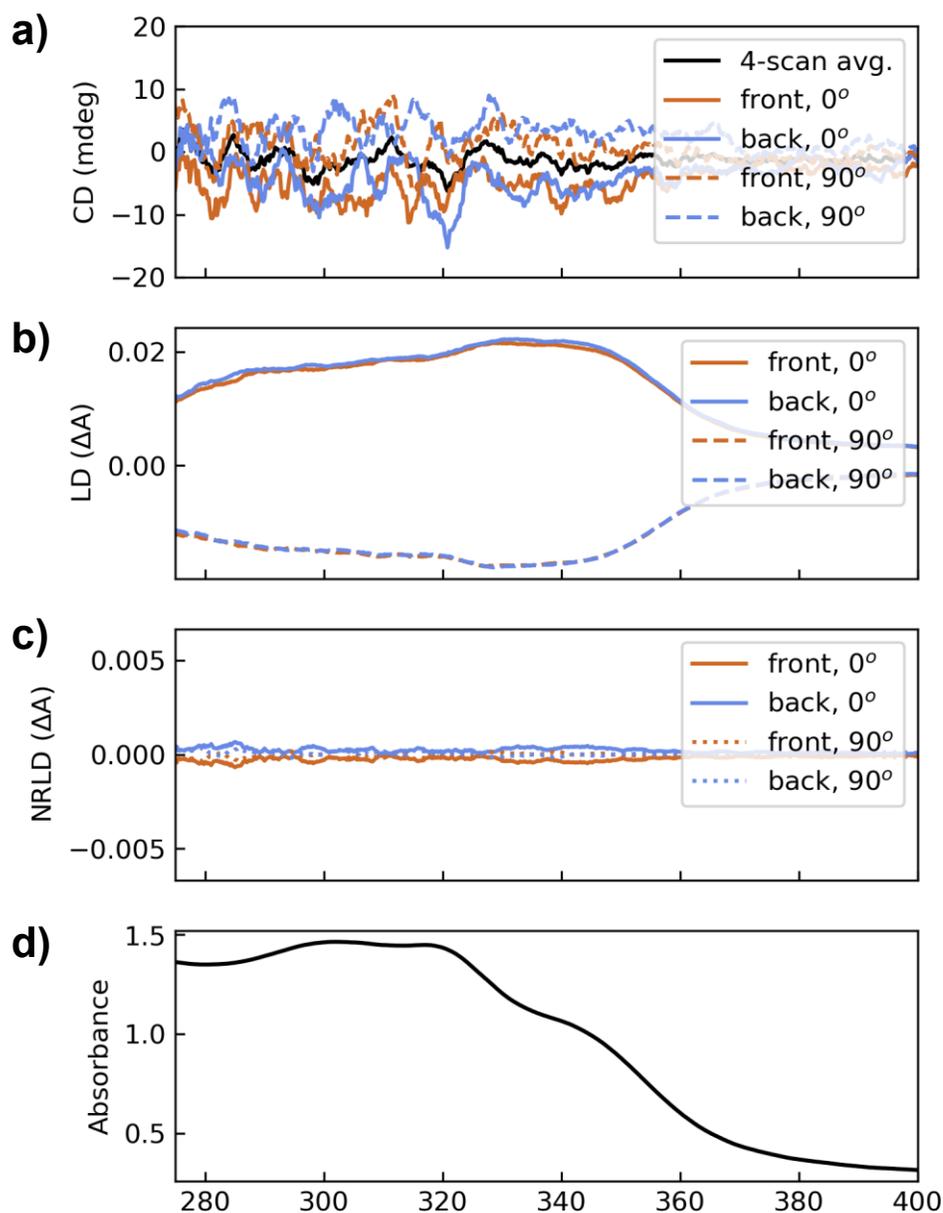

**Figure S12.** Control sample displaying reciprocal LD. A film of linearly-aligned, optically-inactive Cu MSCs was prepared and measured. **a)** CD (solid black curve) measured from a self-assembled Cu MSC film. The black spectra are the average of the four spectra collected for this sample, a methodology derived in our group's previous work (Yao, *et al.*, ACS Nano, 2022). The colored spectra are the individual scans at each orientation. This sample presents no optical activity. **b)** LD spectra collected at each of the four orientations. Because the Cu MSCs are anisotropic, there is a non-zero linear dichroism. **c)** The NRLD components, computed from the spectra shown in panel b, using equations (3) and (4) in the maintext. Due to the lack of optical activity, there is no nonreciprocal linear dichroism component. **d)** Absorbance measured for the Cu MSC film. The absorbance is computed as the average of the four absorbances measured at each orientation.

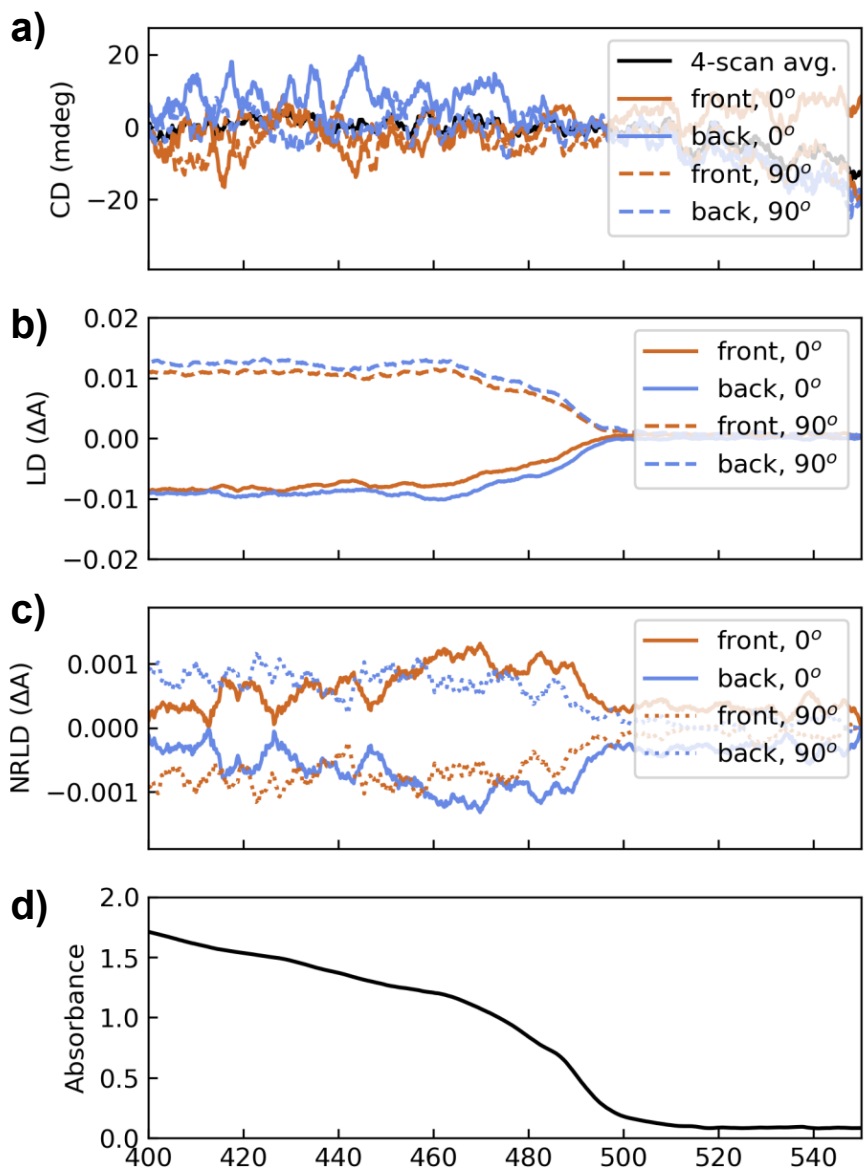

**Figure S13.** Control sample displaying reciprocal LD. A film of linearly-aligned, optically-inactive CdS nanorods was prepared and measured. **a)** CD (solid black curve) measured from a dropcast CdS nanorod film. The black spectra are the average of the four spectra collected for this sample, a methodology derived in our group's previous work (Yao, *et al.*, ACS Nano, 2022). The colored spectra are the individual scans at each orientation. This sample presents no optical activity. **b)** LD spectra collected at each of the four orientations. Because the CdS nanorods are anisotropic, there is a non-zero linear dichroism. **c)** The NRLD components, computed from the spectra shown in panel b, using equations (3) and (4) in the maintext. Due to the lack of optical activity, there is no nonreciprocal linear dichroism component. The small differences between the front and back of the sample are attributed to inhomogeneities in the beam and slight movement of the sample. The NRLD here is more than an order of magnitude weaker, in some cases two orders of magnitude weaker, than what is observed in the optically active CdS, CdSe, and CdTe samples. **d)** Absorbance measured for the CdS nanorod film. The absorbance is computed as the average of the four absorbances measured at each orientation.

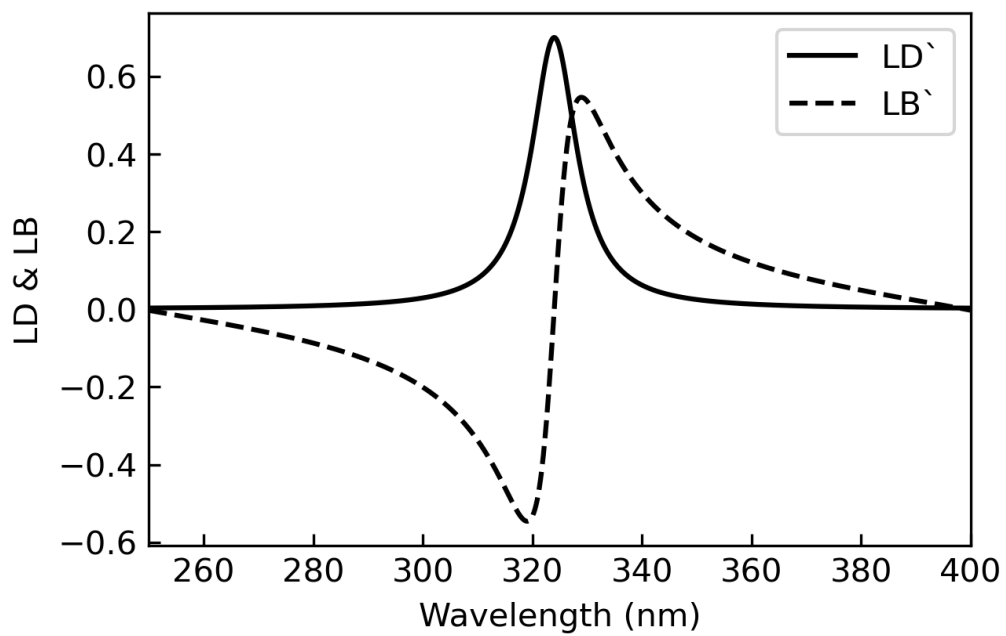

**Figure S14.** LD` and LB` functions used to model the lineshape dependence in **Figure 3c**. LD and LB were both set to 0 for simplicity. The CD spectra used are shown in **Figure 3c**, and the CB was computed from the CB through Kramers-Kronig.

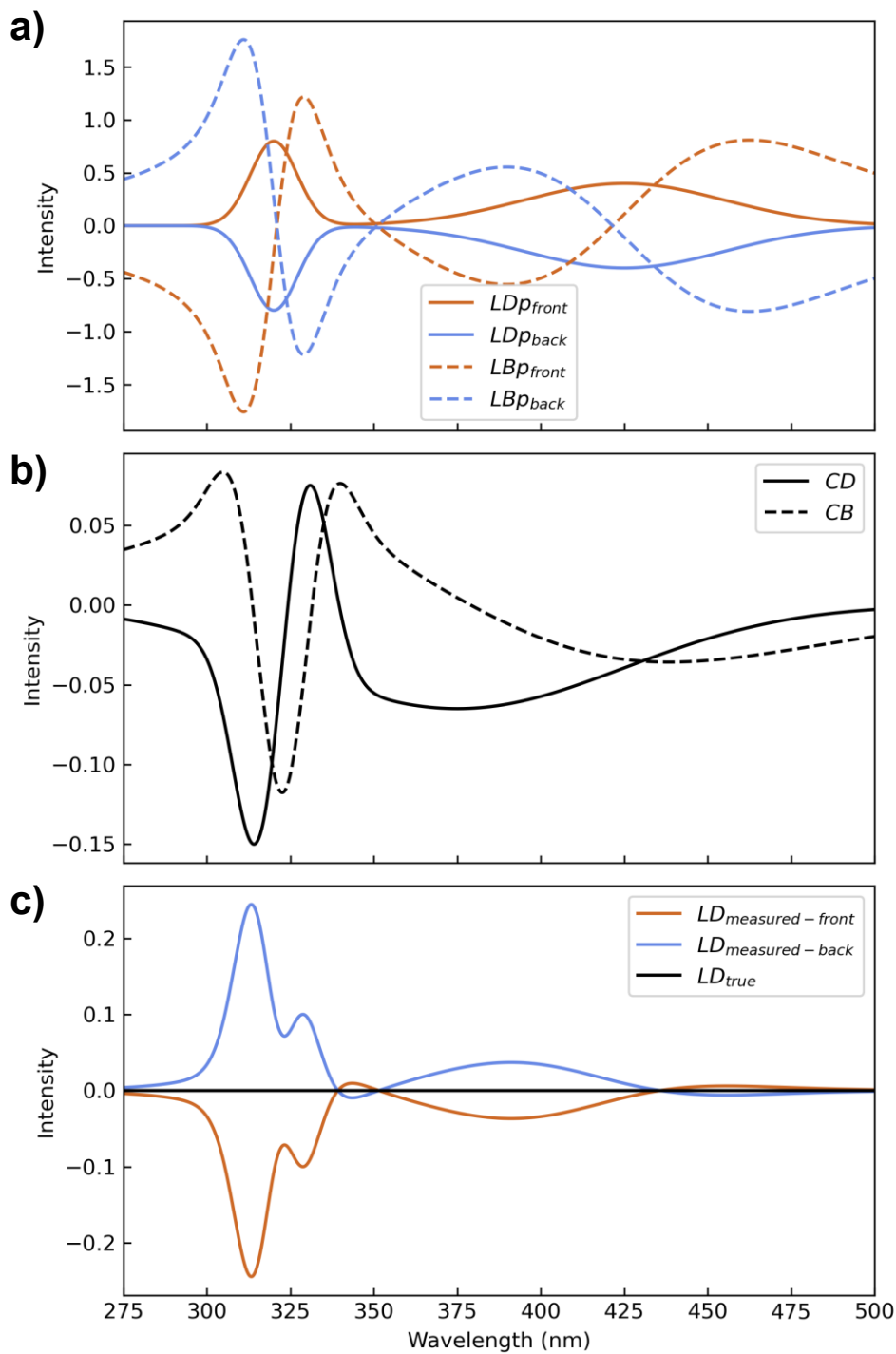

**Figure S15.** Modelled Nonreciprocal LD & LB spectra used in **Figure 3d**. **a)** The LD` and LB` used for the front and back faces of the film. **b)** The CD and CB used for both faces of the film (considered rotationally invariant). **c)** The computed nonreciprocal LD, compared to the "true" LD, which was set to 0. The lineshapes used for the dichroic effects are sums of various Gaussians, used to mimic measured luminescence lineshapes. The birefringent counterparts are computed through Kramers-Kronig.

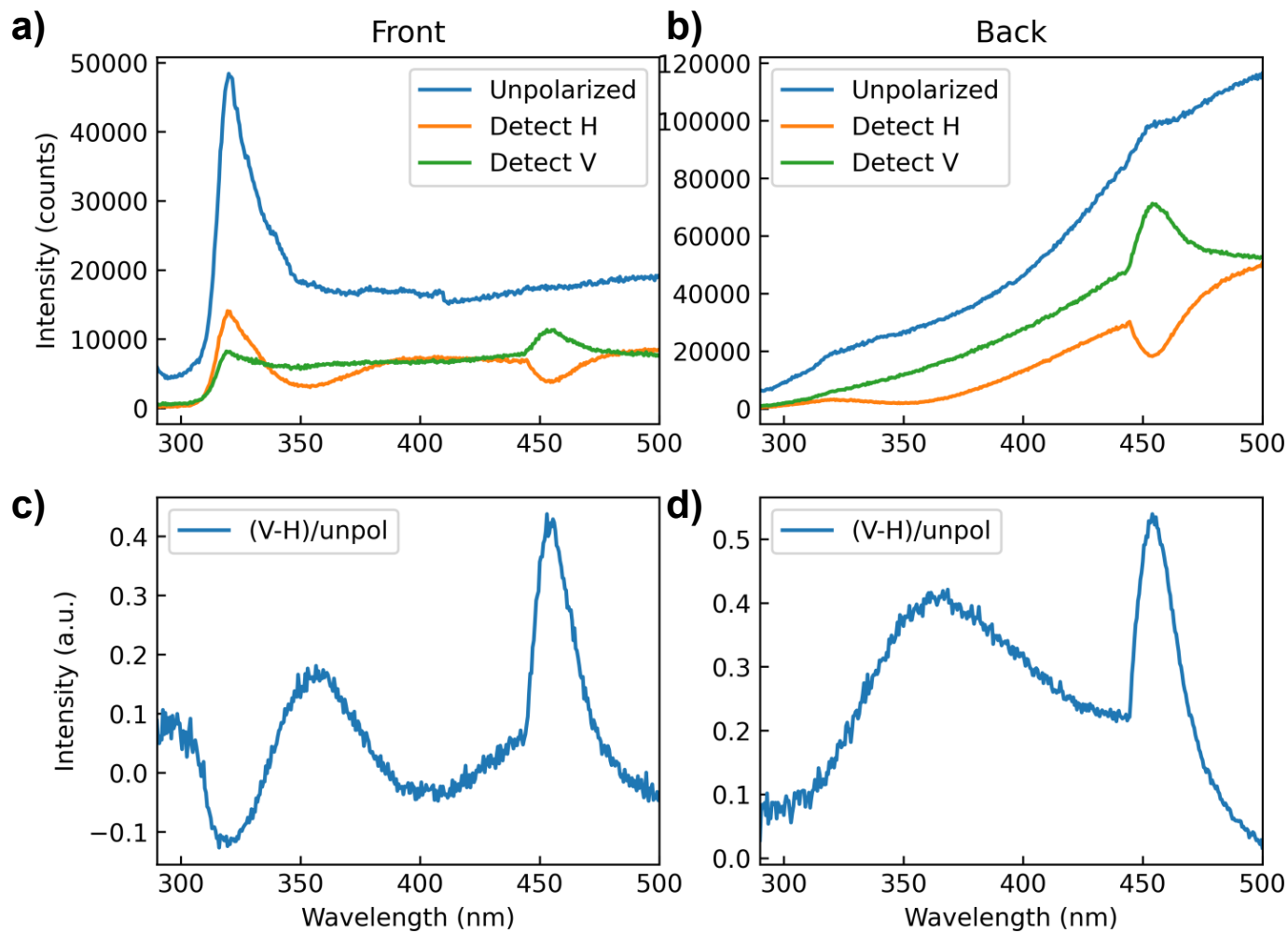

**Figure S16.** LPL experimental data used in **Figure 3d**. **a)** unpolarized and linearly polarized spectra emitted from the front face. **b)** unpolarized and linearly polarized spectra emitted from the back face. **c-d)** difference between the horizontal and vertical emission from the front and back faces, respectively, normalized to the total, unpolarized emission.

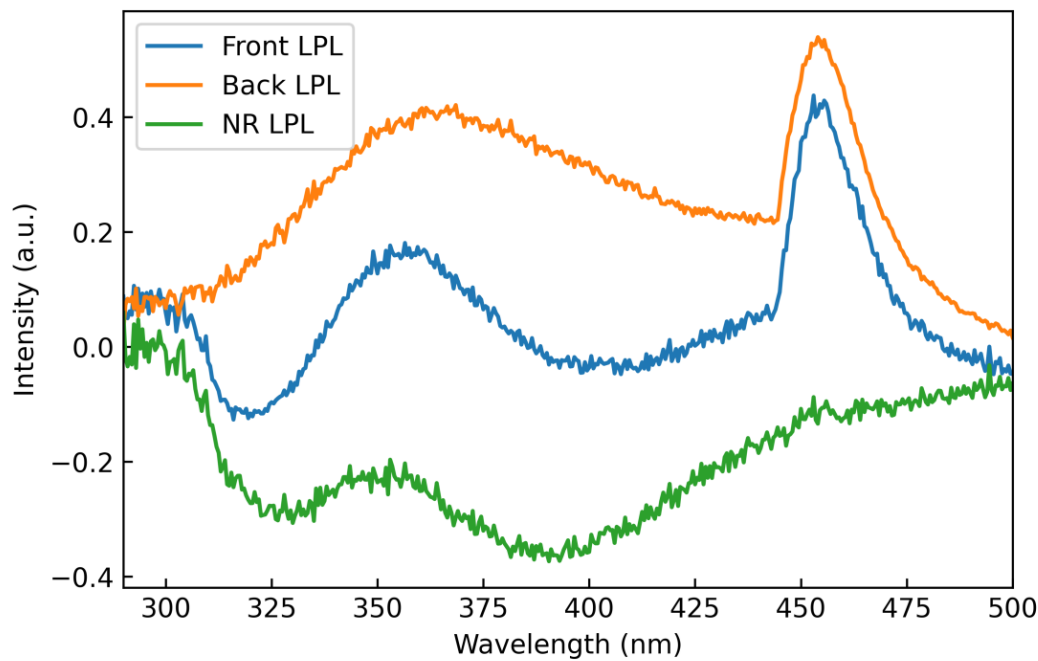

**Figure S17.** LPL experimental data used in **Figure 3d**. The blue and orange spectra are the same as presented in **Fig. S16c-d**, and the green spectrum is the difference between them.

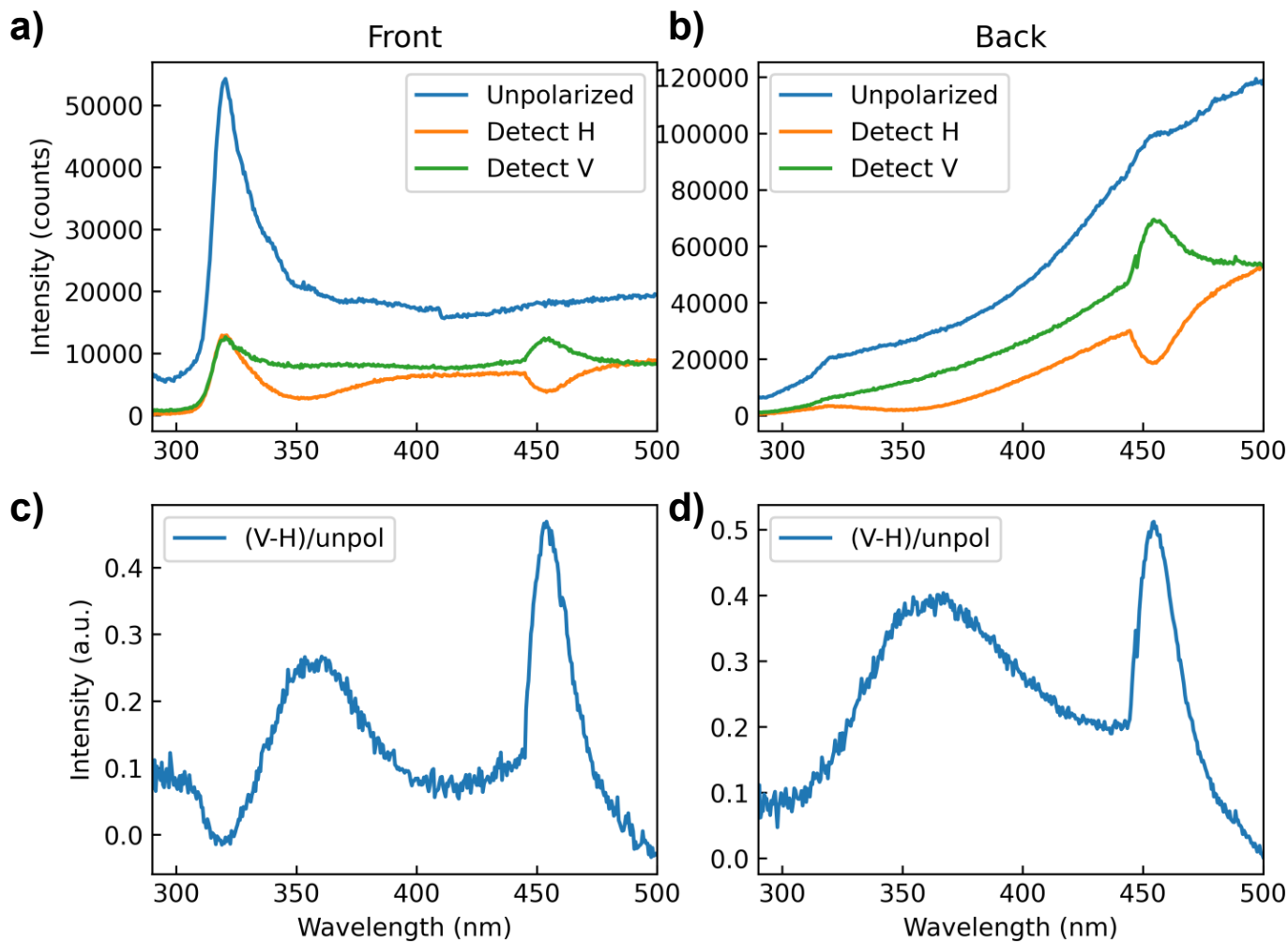

**Figure S18.** Additional LPL experimental data. **a)** unpolarized and linearly polarized spectra emitted from the front face. **b)** unpolarized and linearly polarized spectra emitted from the back face. **c-d)** difference between the horizontal and vertical emission from the front and back faces, respectively, normalized to the total, unpolarized emission.

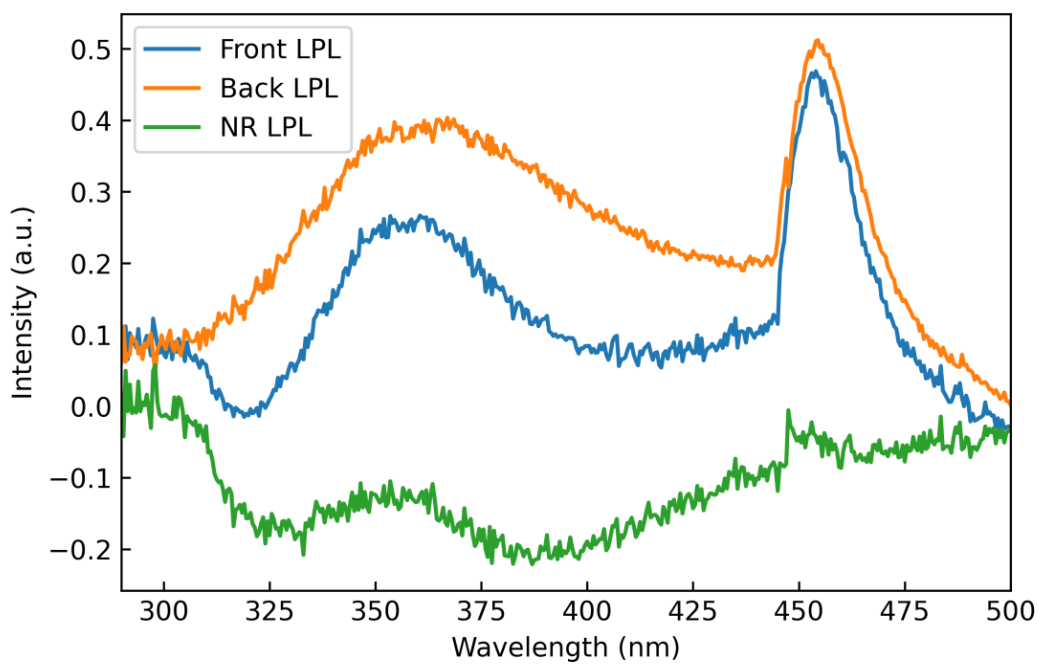

**Figure S19.** Additional LPL experimental data. The blue and orange spectra are the same as presented in **Fig. S18c-d**, and the green spectrum is the difference between them.

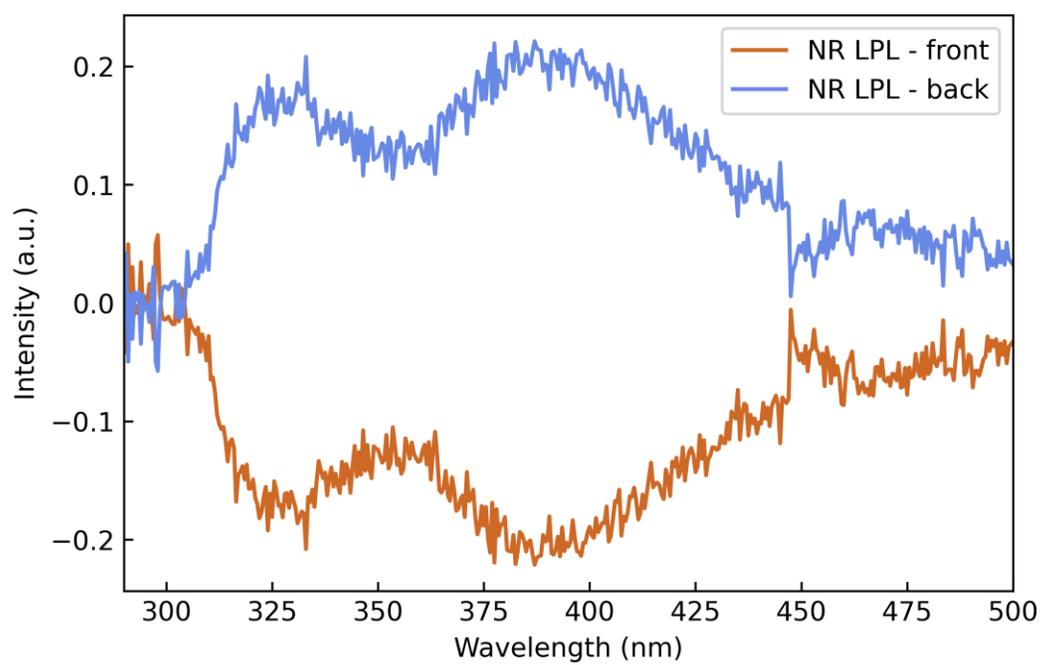

**Figure S20.** Nonreciprocal LPL, isolated from the data presented in **Figs. S18-S19**.

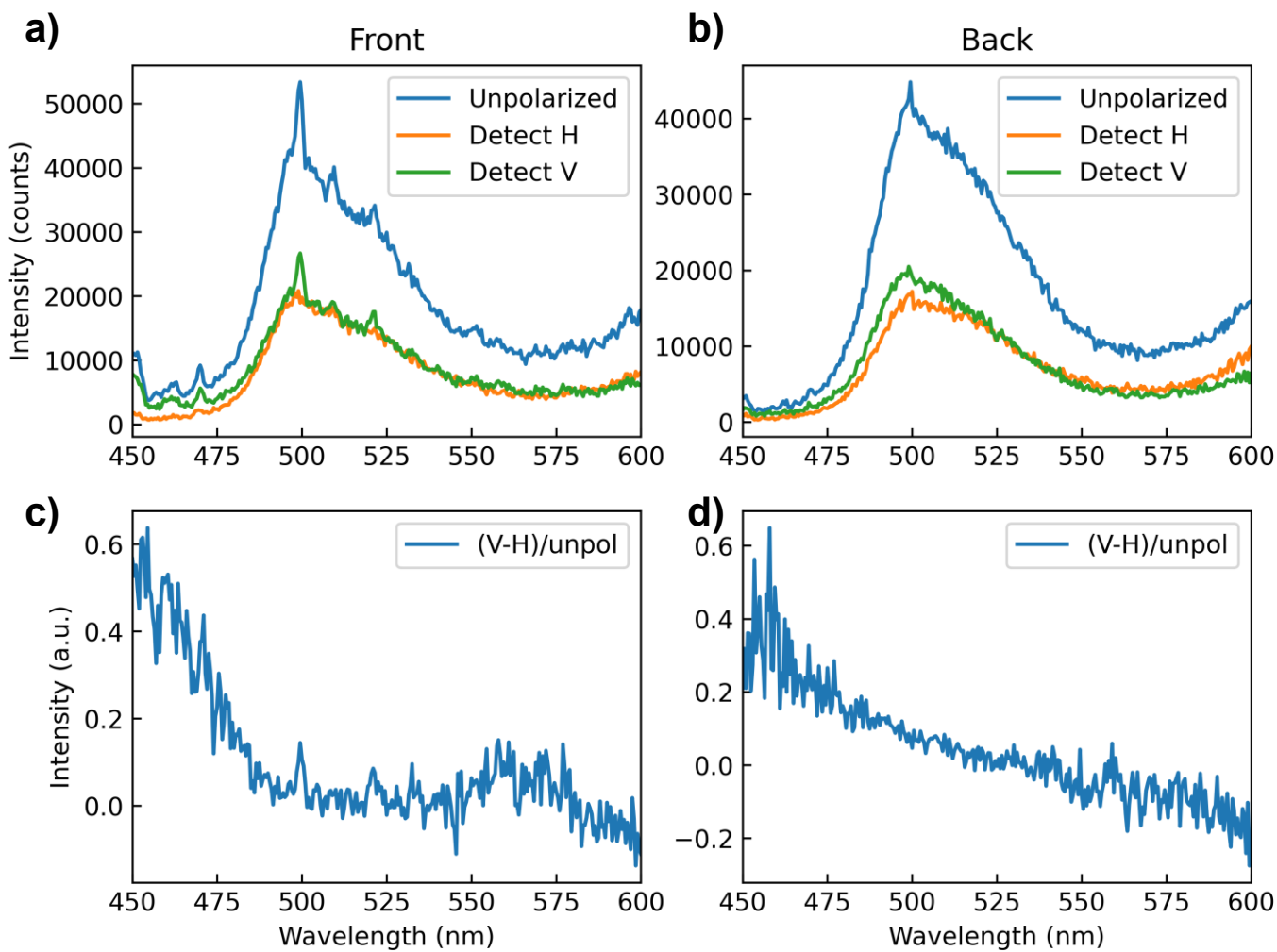

**Figure S21.** Control LPL experimental data from CdS nanorods. **a)** unpolarized and linearly polarized spectra emitted from the front face. **b)** unpolarized and linearly polarized spectra emitted from the back face. **c-d)** difference between the horizontal and vertical emission from the front and back faces, respectively, normalized to the total, unpolarized emission.

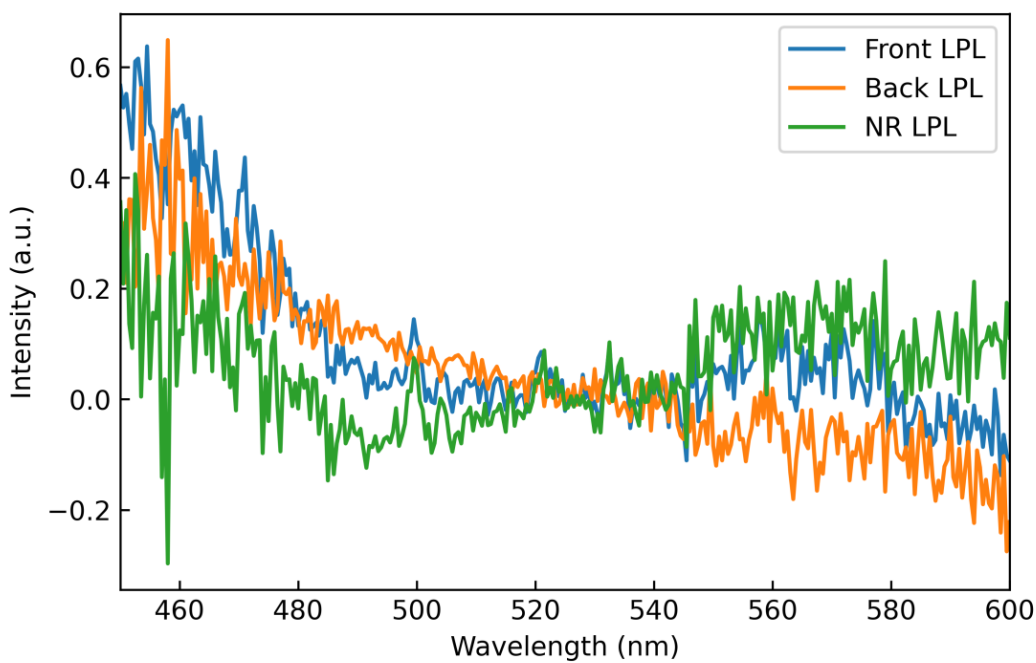

**Figure S22.** Control LPL experimental data from CdS nanorods. The blue and orange spectra are the same as presented in **Fig. S21c-d**, and the green spectrum is the difference between them.

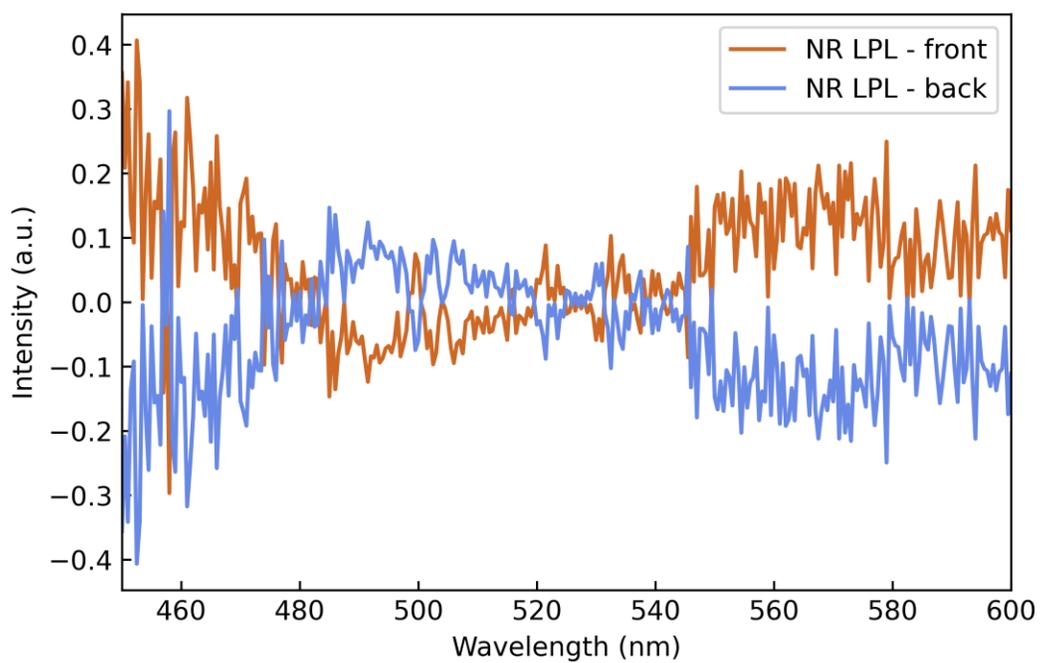

**Figure S23.** Nonreciprocal LPL control from CdS nanorods, isolated from the data presented in **Figs. S21-S22**. Although there are small differences between the spectra, these differences are on the order of the noise levels of the measurement.

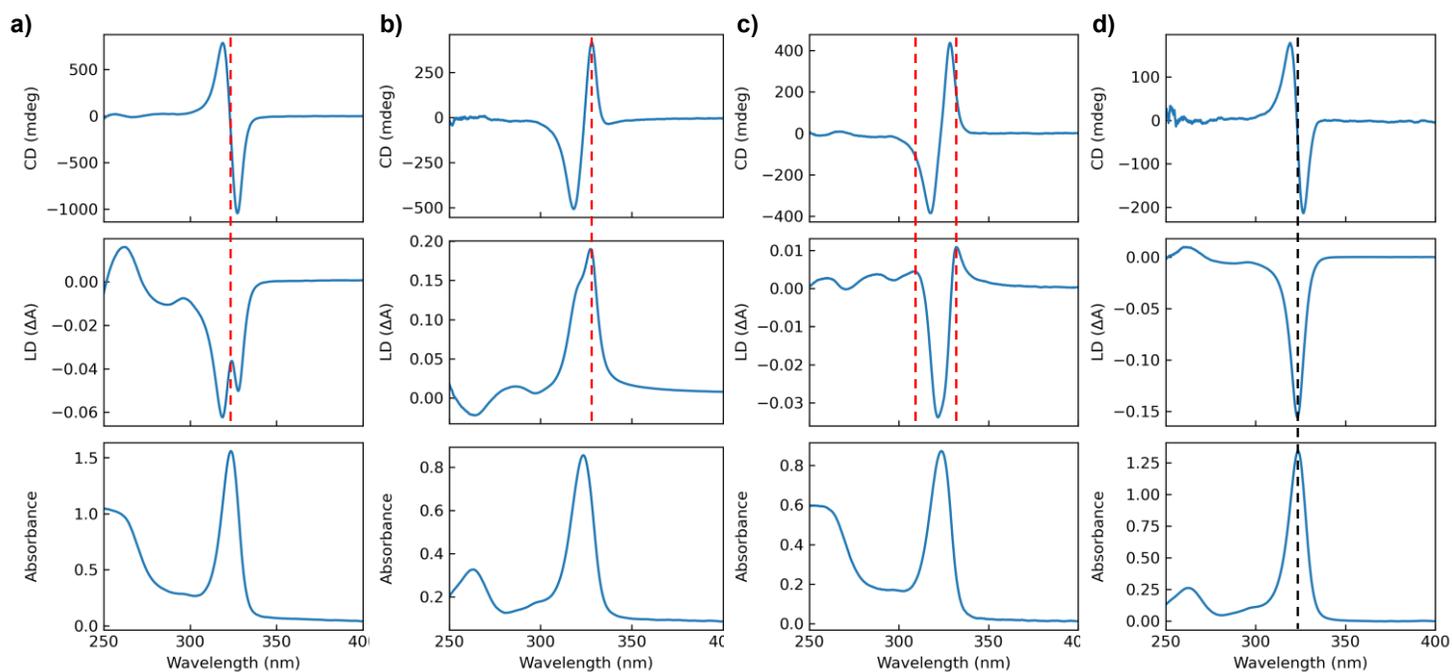

**Figure S24.** Examples of various LD artifacts in MSC films measured on a lab-based CD spectrometer. **a-c)** These measurements present artifactual contributions. The red dashed lines are used to guide the eye to the artifacts. **d)** This measurement does not present significant artifactual contributions.

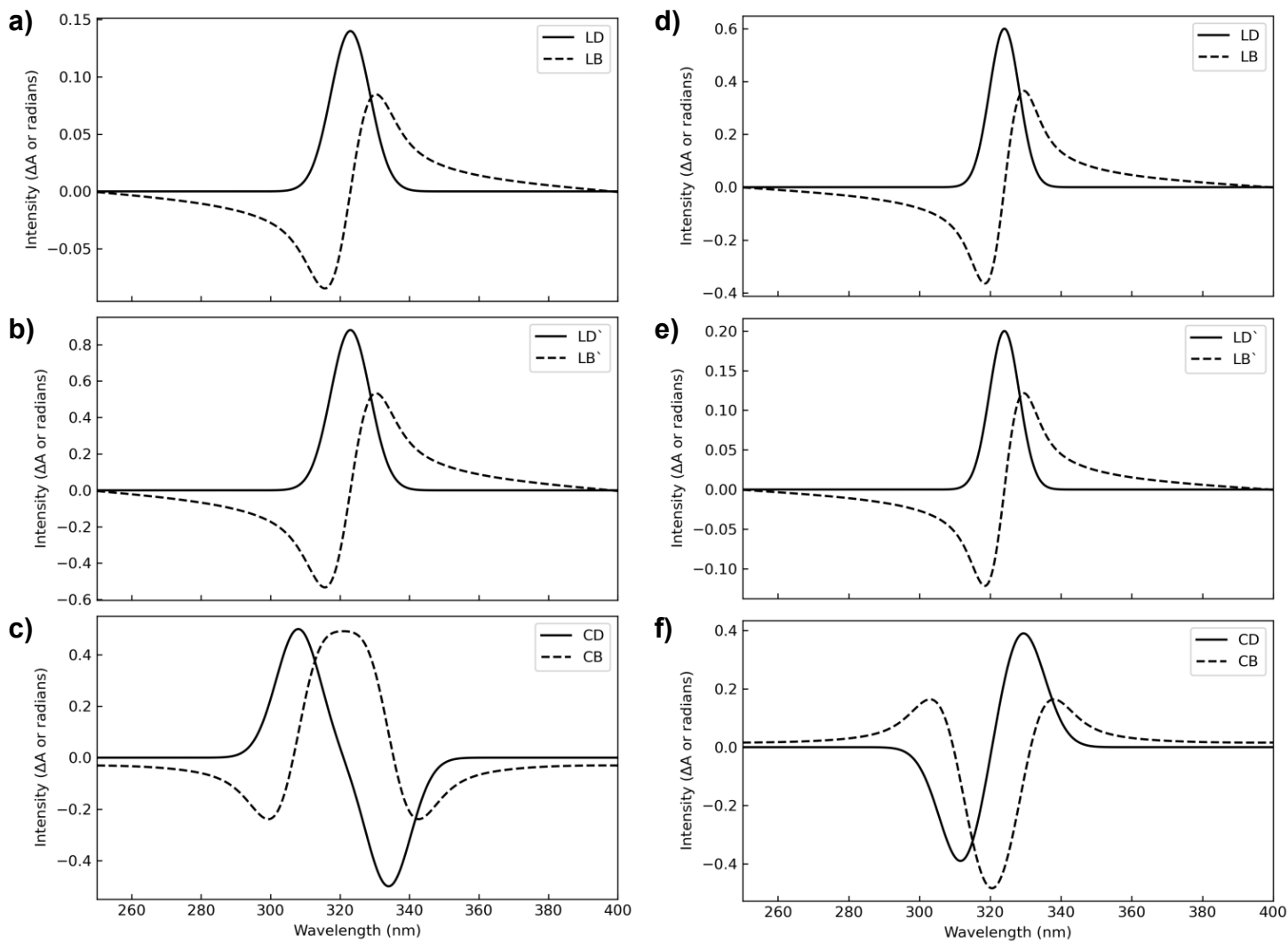

**Figure S25.** Optical effects used to create the plots in **Figure 4**. **a-c)** Dichroic and birefringent parameters used to recreate experimental data in **Figure 4b**. **d-f)** Dichroic and birefringent parameter used to recreate experimental data in **Figure 4c**.

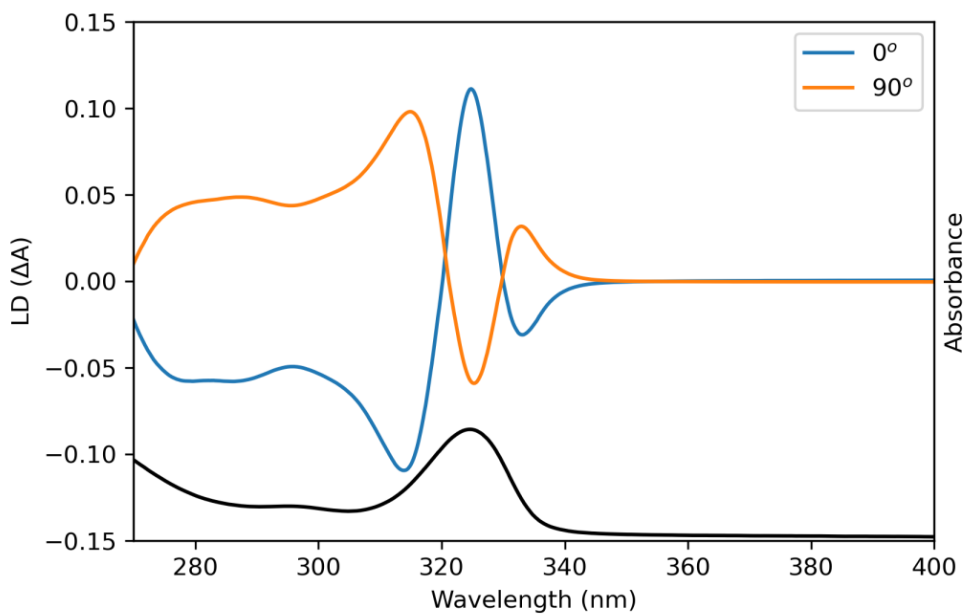

**Figure S26.** Distortions to LD spectra invert upon rotation. This is the dataset used in **Figure 4c**. The artifacts are present & flip sign upon sample rotation. Similarly, the nonreciprocal components invert when a sample is rotated.

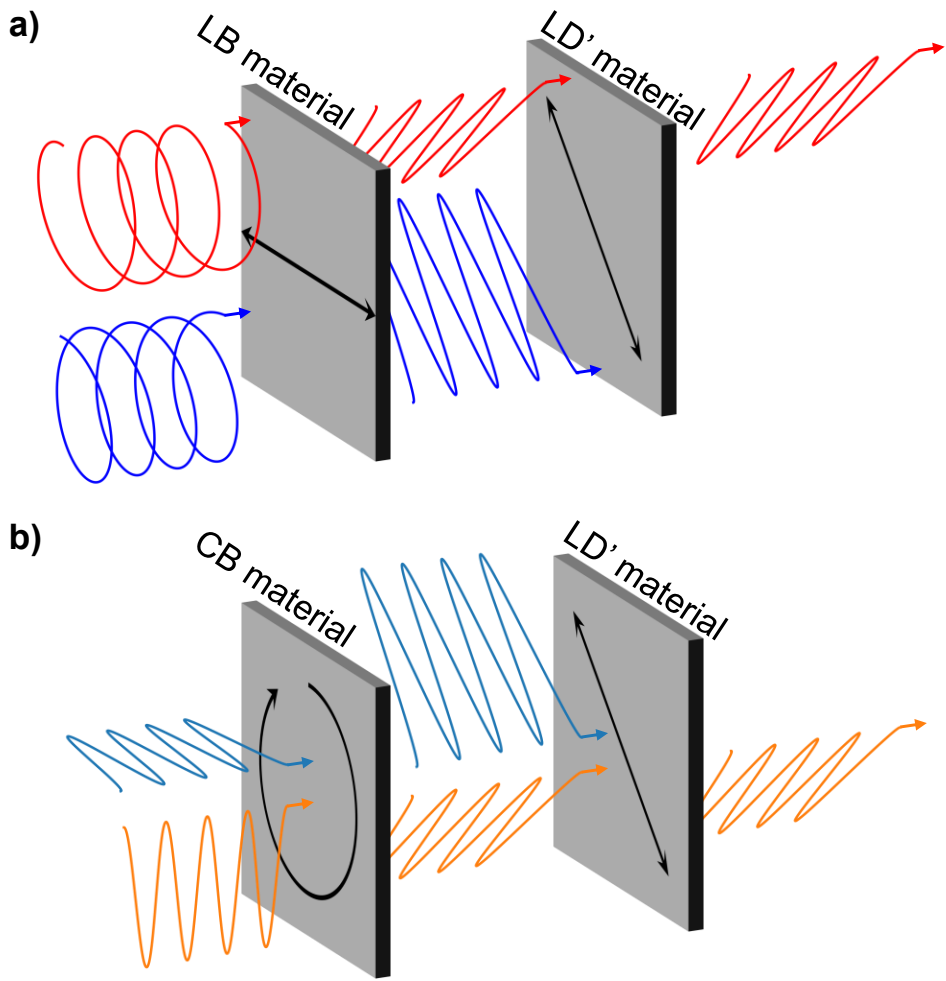

**Figure S27.** Slab model used previously to visualize anisotropic contributions to polarimetry. **a)** Slab picture used to visualize LDLB effects. The circularly-polarized light first transmits through and LB material that projects it to elliptically- or linearly-polarized light, before passing through a LD` material that absorbs the components aligned with the dipoles to yield a CD signal. This picture facilitates visualizing the CD, but it does not portray the nonreciprocal aspect of it. It also is, physically, not the same as progressing through many differential Mueller matrices. **b)** Slab picture that could be used to visualize CBLD` effects. The linearly polarized light progresses through a CB material that rotates them to opposite diagonals. The light then passes through an LD` material, and the polarization aligned with the dipoles is absorbed to a greater degree to yield an LD signal. This picture is helpful for developing an intuition but, much like the above picture, is scientifically inaccurate and does not depict the nonreciprocal aspect.

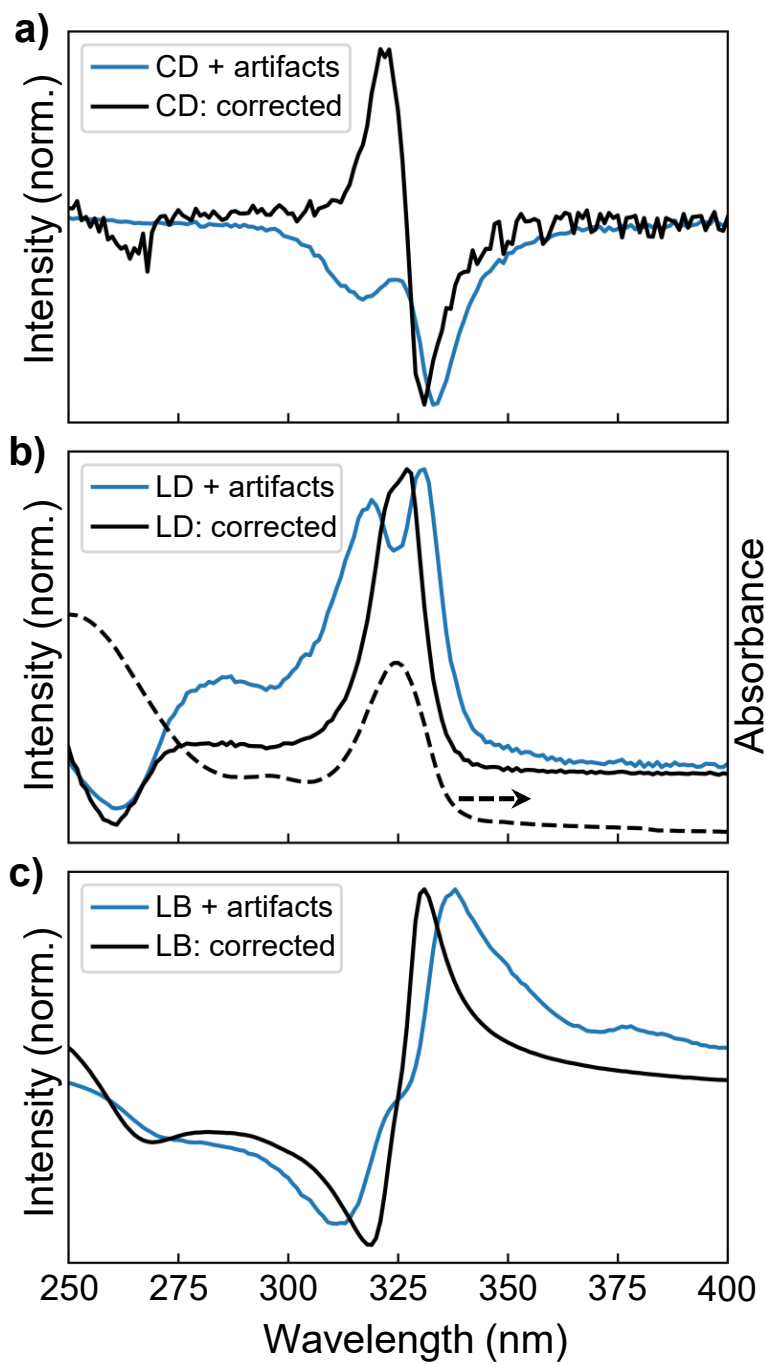

**Figure S28.** Mueller matrix polarimetry of chiral and linear effects in MSC films revealing artifacts in the raw matrix elements, which are accounted for and corrected by analytical inversion. **a)** CD of an MSC film. The raw CD matrix element, $M_{03}$, in blue has a monosignate lineshape with two peaks, caused by both exciton coupling in a chiral assembly and LDLB effects. The CD computed through analytical inversion, in orange, however, has a bisignate lineshape, caused by exciton coupling between MSCs in a chiral assembly. **b)** LD of an MSC film. The raw matrix element, $M_{01}$, in blue has a lineshape with two peaks centered about the absorption maximum, caused by both linear anisotropy within the film and CDLB/CBLD effects. The LD computed through analytical inversion, in orange, however, has a single peak, caused by the linear alignment of MSCs and their transitions within the film. The dashed curve is the excitonic absorbance; the zero-crossing in the CD spectrum occurs at the absorbance peak position. **c)** LB of an MSC film. The raw matrix element, $M_{23}$, in blue has a distorted bisignate lineshape centered at the absorption maximum, caused by both linear anisotropy within the film and CDLB/CBLD effects. The LB computed through analytical inversion, in orange, however, becomes a smooth and undistorted bisignate, caused by the difference in the phase of the light with respect to the natural oscillation frequencies of the electronic states in the clusters.

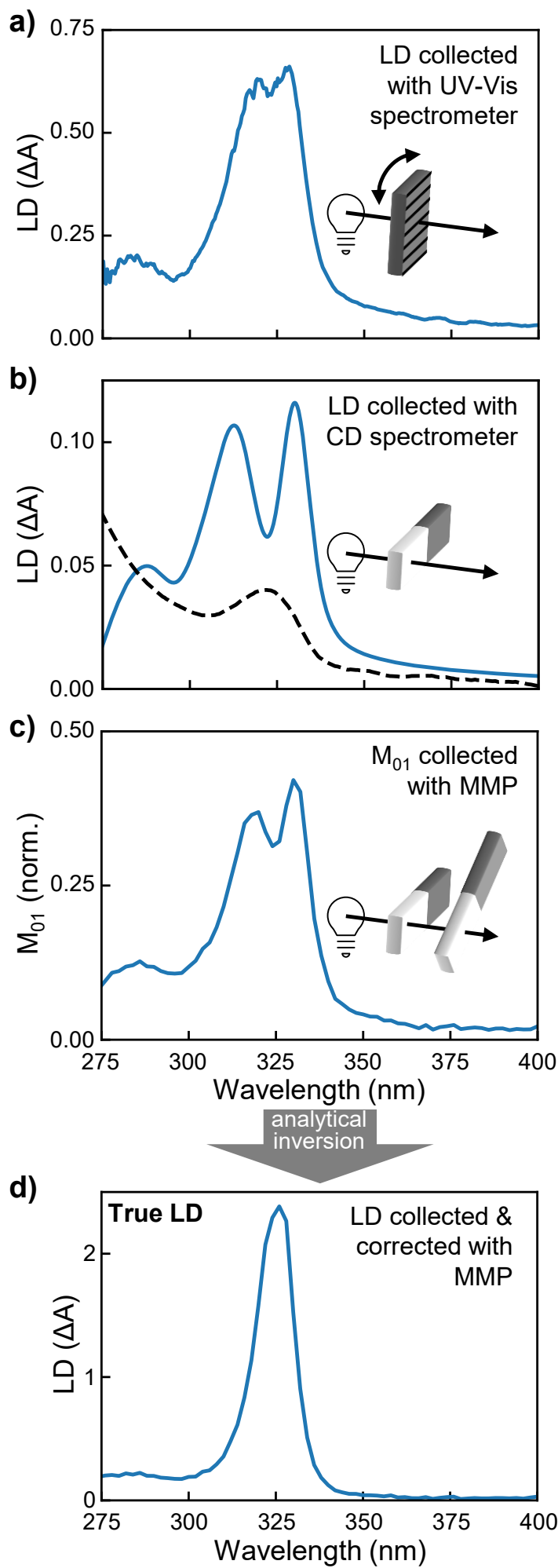

**Figure S29.** Experimental LD measurement of an MSC film resulting in LD distortions. **a-c)** LD spectra measured on a UV-Vis spectrometer, CD spectrometer, and Mueller matrix polarimeter. All measurements yield the same distorted, "doublet" lineshape. **d)** Analytical inversion of Mueller matrix spectra yields an LD lineshape that match the absorption lineshape. Panels **c** and **d** are the same dataset as is used in **Figure 4a**.

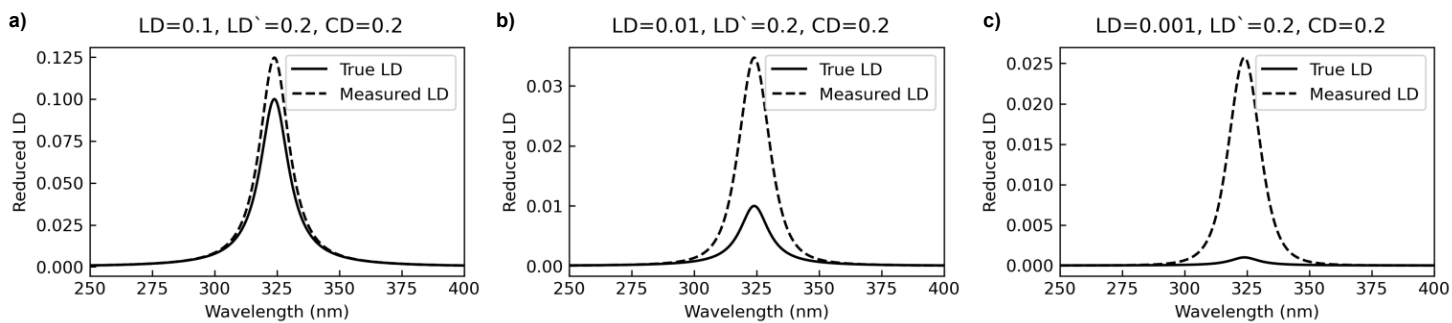

**Figure S30.** Impact of the relative magnitude of the optical effects on the measured LD. **a-c)** True LD spectra vs. Measured LD spectra for three different values of LD, while holding LD` and CD constant.

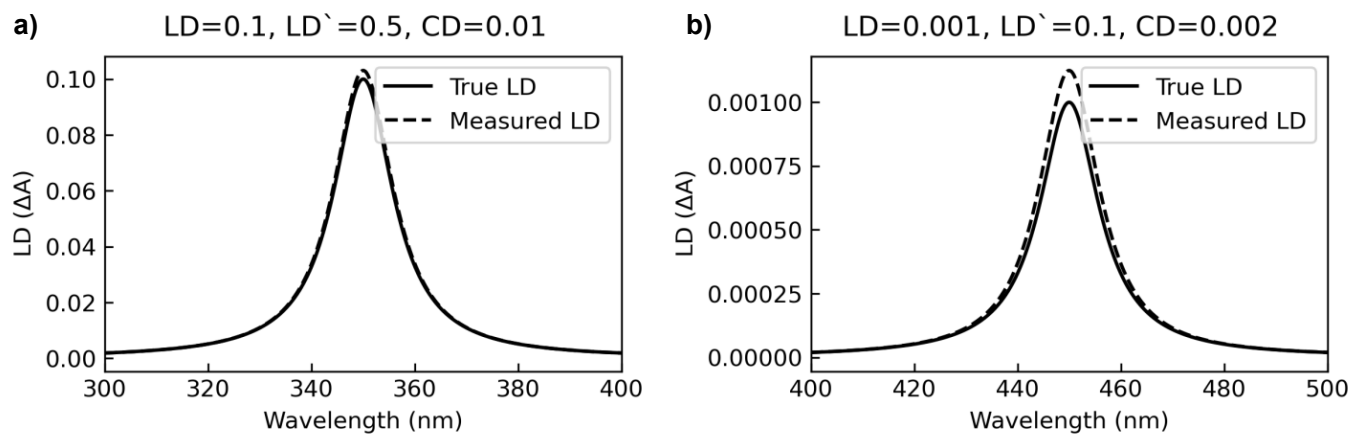

**Figure S31.** Impact of the relative magnitude of the optical effects on the measured LD for CdSe and CdTe. **a)** Modelled True LD spectra vs. Measured LD spectra for the values of CdSe optical effects. **b)** Modelled True LD spectra vs. Measured LD spectra for the values of CdTe optical effects.